\documentclass[25pt]{article}
\usepackage{amsfonts,amsmath,amssymb,amsthm,epsfig,mathrsfs,subfigure}
\usepackage{setspace}
\theoremstyle{definition}

\theoremstyle{plain}

\title{Riemann-Hilbert approach and $N$-soliton solution for an eighth-order nonlinear Schr\"{o}dinger equation in an optical fiber}
\author{Zhou-Zheng Kang$^{1,2}$, Tie-Cheng Xia$^{1}$\footnote{Corresponding author. E-mail: xiatc@shu.edu.cn.} \\
1. Department of Mathematics, Shanghai University, Shanghai 200444, China;\\
2. College of Mathematics, Inner Mongolia University for Nationalities,\\ Tongliao 028043, China\\}
\date{}
\begin{document}
\sloppy \maketitle
\begin{abstract}
ժҪThis paper aims to present an application of Riemann-Hilbert approach to treat higher-order nonlinear differential equation that is an eighth-order nonlinear Schr\"{o}dinger equation arising in an optical fiber.
Starting from the spectral analysis of the Lax pair, a Riemann-Hilbert problem is formulated. Then by solving the obtained Riemann-Hilbert problem under the reflectionless case, $N$-soliton solution is generated for the eighth-order nonlinear Schr\"{o}dinger equation. Finally, the three-dimensional plots and two-dimensional curves with specific choices of the involved parameters are made to show the localized structures and dynamic behaviors of one- and two-soliton solutions. \vskip
2mm\noindent\textbf{AMS Subject classification}: 35C08\vskip
2mm\noindent\textbf{Keywords}: eighth-order nonlinear Schr\"{o}dinger equation; Riemann-Hilbert approach; soliton solutions
\end{abstract}
\section{Introduction}
In this paper, we investigate in detail an eighth-order nonlinear Schr\"{o}dinger (NLS) equation
\begin{equation}
\begin{aligned}
&i{{q}_{t}}+{{A}_{2}}{{K}_{2}}[q(x,t)]-i{{A}_{3}}{{K}_{3}}[q(x,t)]+{{A}_{4}}{{K}_{4}}[q(x,t)]-i{{A}_{5}}{{K}_{5}}[q(x,t)]\\
&+{{A}_{6}}{{K}_{6}}[q(x,t)]-i{{A}_{7}}{{K}_{7}}[q(x,t)]+{{A}_{8}}{{K}_{8}}[q(x,t)]=0,
\end{aligned}
\end{equation}
which is used for describing the propagation of ultrashort nonlinear pulses.$^{1,2}$ It can be generated from truncating the infinite hierarchy of nonlinear Schr\"{o}dinger equations$^{3}$ that is used to investigate the higher-order dispersive effects and nonlinearity.
Here $q(x,t)$ denotes a normalized complex amplitude of the optical pulse envelope. And the subscripts of $q(x,t)$ mean the partial derivatives with respect to the scaled spatial coordinate $x$ and time coordinate $t$ correspondingly. Each coefficient $A_{j}(2\leq j\leq 8)$ is an arbitrary real number, and
\begin{equation*}
\begin{aligned}
&{{K}_{2}}[q(x,t)]={{q}_{xx}}+2q{{\left| q \right|}^{2}},\\
&{{K}_{3}}[q(x,t)]={{q}_{xxx}}+6{{\left| q \right|}^{2}}{{q}_{x}},\\
&{{K}_{4}}[q(x,t)]={{q}_{xxxx}}+6{{q}^{*}}q_{x}^{2}+4q{{\left| {{q}_{x}} \right|}^{2}}+8{{\left| q \right|}^{2}}{{q}_{xx}}+2{{q}^{2}}q_{xx}^{*}+6{{\left| q \right|}^{4}}q,\\
&{{K}_{5}}[q(x,t)]={{q}_{xxxxx}}+10{{\left| q \right|}^{2}}{{q}_{xxx}}+30{{\left| q \right|}^{4}}{{q}_{x}}+10q{{q}_{x}}q_{xx}^{*}+10qq_{x}^{*}{{q}_{xx}}+20{{q}^{*}}{{q}_{x}}{{q}_{xx}}+10q_{x}^{2}q_{x}^{*},\\
&{{K}_{6}}[q(x,t)]={{q}_{xxxxxx}}+{{q}^{2}}\big[60{{\left| {{q}_{x}} \right|}^{2}}{{q}^{*}}+50{{q}_{xx}}{{({{q}^{*}})}^{2}}+2q_{xxxx}^{*}\big]+q\big[12{{q}_{xxxx}}{{q}^{*}}+8{{q}_{x}}q_{xxx}^{*}\\&\quad\quad\quad\quad\quad\ \ +22{{\left| {{q}_{xx}}\right|}^{2}}+18{{q}_{xxx}}q_{x}^{*}+70q_{x}^{2}{{({{q}^{*}})}^{2}}\big]+20q_{x}^{2}q_{xx}^{*} +10{{q}_{x}}\big(5{{q}_{xx}}q_{x}^{*}+3{{q}_{xxx}}{{q}^{*}}\big)\\&\quad\quad\quad\quad\quad\ \  +20q_{xx}^{2}{{q}^{*}}+10{{q}^{3}}\big[{{(q_{x}^{*})}^{2}}+2{{q}^{*}}q_{xx}^{*}\big]+20q{{\left| q \right|}^{6}},\\
&{{K}_{7}}[q(x,t)]={{q}_{xxxxxxx}}+70q_{xx}^{2}q_{x}^{*}+112{{q}_{x}}{{\left| {{q}_{xx}} \right|}^{2}}+98{{q}_{xxx}}{{\left| {{q}_{x}} \right|}^{2}}+70{{q}^{2}}\big[{{q}_{x}}{{(q_{x}^{*})}^{2}} +2{{q}_{x}}{{q}^{*}}q_{xx}^{*}\\&\quad\quad\quad\quad\quad\ \ +{{q}^{*}}\big(2{{q}_{xx}}q_{x}^{*} +{{q}_{xxx}}{{q}^{*}}\big)\big]+28q_{x}^{2}q_{xxx}^{*}+14q\big[{{q}^{*}}\big(20{{\left| {{q}_{x}} \right|}^{2}}{{q}_{x}} +{{q}_{xxxxx}}\big)+3{{q}_{xxx}}q_{xx}^{*}\\&\quad\quad\quad\quad\quad\ \  +2{{q}_{xx}}q_{xxx}^{*}+2{{q}_{x}}q_{xxxx}^{*}+20{{q}_{x}}{{q}_{xx}}{{({{q}^{*}})}^{2}}\big]+140{{\left| q \right|}^{6}}{{q}_{x}}+70q_{x}^{3}{{({{q}^{*}})}^{2}}\\&\quad\quad\quad\quad\quad\ \ +14{{q}^{*}}(5{{q}_{xx}}{{q}_{xxx}}+3{{q}_{x}}{{q}_{xxxx}}).\\
&{{K}_{8}}[q(x,t)]={{q}_{xxxxxxxx}}+14{{q}^{3}}\big[40{{\left| {{q}_{x}} \right|}^{2}}{{({{q}^{*}})}^{2}}+20{{q}_{xx}}{{({{q}^{*}})}^{3}}+2q_{xxxx}^{*}{{q}^{*}}+3{{(q_{xx}^{*})}^{2}}
+4q_{x}^{*}q_{xxx}^{*}\big]\\&\quad\quad\quad\quad\quad\ \ +{{q}^{2}}\big[28{{q}^{*}}(14{{q}_{xx}}q_{xx}^{*}
+11{{q}_{xxx}}q_{x}^{*}+6{{q}_{x}}q_{xxx}^{*})
+238{{q}_{xx}}{{(q_{x}^{*})}^{2}}+336{{\left| {{q}_{x}} \right|}^{2}}q_{xx}^{*}\\&\quad\quad\quad\quad\quad\ \ +560q_{x}^{2}{{({{q}^{*}})}^{3}}+98{{q}_{xxxx}}{{({{q}^{*}})}^{2}}
+2q_{xxxxxx}^{*}\big]
 +2q\big\{21q_{x}^{2}\big[9{{(q_{x}^{*})}^{2}}+14{{q}^{*}}q_{xx}^{*}\big]\\&\quad\quad\quad\quad\quad\ \ +{{q}_{x}}\big[728{{q}_{xx}}q_{x}^{*}{{q}^{*}}
 +238{{q}_{xxx}}{{({{q}^{*}})}^{2}}+6q_{xxxxx}^{*}\big]+34{{\left| {{q}_{xxx}} \right|}^{2}}+36{{q}_{xxxx}}q_{xx}^{*}\\&\quad\quad\quad\quad\quad\ \ +22{{q}_{xx}}q_{xxxx}^{*}
 +20{{q}_{xxxxx}}q_{x}^{*}+161q_{xx}^{2}{{({{q}^{*}})}^{2}}+8{{q}_{xxxxxx}}{{q}^{*}}\big\}+182{{q}_{xx}}{{\left| {{q}_{xx}} \right|}^{2}}\\&\quad\quad\quad\quad\quad\ \
 +308{{q}_{xx}}{{q}_{xxx}}q_{x}^{*}+252{{q}_{x}}{{q}_{xxx}}q_{xx}^{*} +196{{q}_{x}}{{q}_{xx}}q_{xxx}^{*}+168{{q}_{x}}{{q}_{xxxx}}q_{x}^{*}\\&\quad\quad\quad\quad\quad\ \
 +42q_{x}^{2}q_{xxxx}^{*}+14{{q}^{*}}\big(30q_{x}^{3}q_{x}^{*}+4{{q}_{xxxxx}}{{q}_{x}}+5q_{xxx}^{2}+8{{q}_{xx}}{{q}_{xxxx}}\big) \\&\quad\quad\quad\quad\quad\ \
 +490q_{x}^{2}{{q}_{xx}}{{({{q}^{*}})}^{2}}+140{{q}^{4}}{{q}^{*}}\big[{{(q_{x}^{*})}^{2}}+{{q}^{*}}q_{xx}^{*}\big]+7q{{\left| q \right|}^{8}}.
\end{aligned}
\end{equation*}
Here the superscript $\ast$ represents complex conjugate.

As a matter of fact, Equation (1) covers many nonlinear differential equations of important significance, some of which are listed as follows:

(i) For the case of $A_{3}=A_{4}=A_{5}=A_{6}=A_{7}=A_{8}=0$, Equation (1) is reduced to the fundamental nonlinear Schr\"{o}dinger equation describing the propagation of the picosecond pulses in an optical fiber.

(ii) For the case of $A_{2}=\frac{1}{2}$ and $A_{4}=A_{5}=A_{6}=A_{7}=A_{8}=0$, Equation (1) is reduced to  the Hirota equation$^{4,5}$ describing the third-order dispersion and time-delay correction to the cubic nonlinearity in ocean waves.

(iii) For the case of $A_{2}=\frac{1}{2}$ and $A_{5}=A_{6}=A_{7}=A_{8}=0$, Equation (1) becomes a fourth-order dispersive NLS equation$^{6,7}$ describing the ultrashort optical-pulse propagation in a long-distance, high-speed optical fiber transmission system.

(iv) For the case of $A_{2}=\frac{1}{2}$ and $A_{6}=A_{7}=A_{8}=0$, Equation (1) becomes a fifth-order NLS equation$^{8}$ describing the attosecond pulses in an optical fiber.

By now, there have been plenty of researches on Equation (1). For instance,
the interactions among multiple solitons were under study,$^{1}$ and oscillations in the interaction zones were observed systematically. As a result, it was found that the oscillations in the solitonic interaction zones possess different forms with different spectral parameters of Equation (1) and so forth.
In a follow-up study,$^{2}$ the Lax pair and infinitely-many conservation laws were derived via symbolic computation, which verifies
the integrability of Equation (1). Moreover, the one-, two- and three-soliton solutions were explored as well by means of the Darboux transformation.

The principal aim of this study is to determine multi-soliton solutions for the eighth-order NLS equation (1) with the aid of the Riemann-Hilbert approach.$^{9-20}$ This paper is divided into five sections. In second section, we recall the Lax pair associated with Equation (1) and convert it into a desired form. In third section,
we carry out the spectral analysis, from which a Riemann-Hilbert problem is set up on the real line. In fourth section, the construction of multi-soliton solutions for Equation (1) is detailedly discussed in the framework of the Riemann-Hilbert problem under the reflectionless case. A brief conclusion is given in the final section.

\section{Lax pair}

Upon the Ablowitz-Kaup-Newell-Segur formalism, the eighth-order NLS equation (1) admits a $2\times2$ Lax pair$^{2}$
\begin{subequations}
\begin{align}
  & {{\Psi}_{x}}=U\Psi ,\quad U=i\left( \begin{matrix}
   \varsigma  & {q}^{*}  \\
   q & -\varsigma   \\
\end{matrix} \right), \\
 & {{\Psi }_{t}}=V\Psi ,\quad V=\sum\limits_{j=0}^{8}{{{\varsigma}^{j}}\left( \begin{matrix}
   {{a}_{j}} & {{b}_{j}}  \\
   {{c}_{j}} & -{{a}_{j}}  \\
\end{matrix} \right)},
\end{align}
\end{subequations}
where $\Psi ={({{\Psi }_{1}},{{\Psi }_{2}})^\textrm{T}}$ is a vector eigenfunction, ${{\Psi }_{1}}$ and ${{\Psi }_{2}}$ are the complex functions of $x$ and $t$, the symbol $\textrm{T}$ signifies transpose of the vector, and $\varsigma$ is a isospectral parameter. Furthermore,

\[\begin{aligned}& {{a}_{0}}=-i{{A}_{8}}\big\{35{{\left| q \right|}^{8}}+21q_{xx}^{2}{{({{q}^{*}})}^{2}}-21{{\left| {{q}_{x}} \right|}^{4}}+14{{q}^{*}}q_{xx}^{*}+70{{q}^{3}}{{q}^{*}}({{(q_{x}^{*})}^{2}}+{{q}^{*}}q_{xx}^{*})\\&\quad\quad-{{q}_{xxx}}q_{xxx}^{*}+q_{xx}^{*}{{q}_{xxxx}}+7{{q}^{2}}\big[2{{q}^{*}}q_{xxxx}^{*}+3{{(q_{xx}^{*})}^{2}}+4q_{x}^{*}q_{xxx}^{*}+10{{q}_{x}}{{({{q}^{*}})}^{2}}q_{x}^{*}\\&\quad\quad+10{{q}_{xx}}{{({{q}^{*}})}^{3}}\big]-{{q}_{xxxxx}}q_{x}^{*}+{{q}_{x}}(28{{q}^{*}}q_{x}^{*}{{q}_{xx}} +28{{({{q}^{*}})}^{2}}{{q}_{xxx}}-q_{xxxxx}^{*})\\&\quad\quad+{{q}^{*}}{{q}_{xxxxxx}}+q\big[70{{({{q}^{*}})}^{3}}{{({{q}_{x}})}^{2}}+14{{(q_{x}^{*})}^{2}}{{q}_{xx}}+28{{q}_{x}}q_{x}^{*}q_{xx}^{*}+14q_{x}^{*}(4{{q}_{xx}}q_{xx}^{*} \\&\quad\quad+q_{x}^{*}{{q}_{xxx}}+{{q}_{x}}q_{xxx}^{*})+14{{({{q}^{*}})}^{2}}{{q}_{xxxx}}+q_{xxxxxx}^{*}\big\}+{{A}_{7}}\big\{-30{{q}^{3}}{{({{q}^{*}})}^{2}}q_{x}^{*}
 \\&\quad\quad+20{{({{q}^{*}})}^{2}}{{q}_{x}}{{q}_{xx}}+q_{xx}^{*}{{q}_{xxx}} +10{{q}^{2}}\big[3{{({{q}^{*}})}^{3}}{{q}_{x}}-2q_{x}^{*}q_{xx}^{*}-{{q}^{*}}q_{xxx}^{*}\big]-{{q}_{xx}}q_{xxx}^{*}\\&\quad\quad-q_{x}^{*}{{q}_{xxxx}}+{{q}_{x}}q_{xxxx}^{*}+{{q}^{*}}(10q_{x}^{2}q_{x}^{*}+{{q}_{xxxxx}}) -q\big[10{{q}^{*}}q_{x}^{*}{{q}_{xx}}+10{{q}_{x}}({{(q_{x}^{*})}^{2}}\\&\quad\quad-{{q}^{*}}q_{xx}^{*})\big]-10{{({{q}^{*}})}^{2}}{{q}_{xxx}}+q_{xxxxx}^{*}\big\}-i{{A}_{6}}\big\{10{{\left| q \right|}^{6}}+5{{({{q}^{*}})}^{2}}q_{xx}^{2}+{{q}_{xx}}q_{xx}^{*}\\&\quad\quad+5{{q}^{2}}\big[{{(q_{x}^{*})}^{2}}+2{{q}^{*}}q_{xx}^{*}\big]-q_{x}^{*}{{q}_{xxx}}-{{q}_{x}}q_{xxx}^{*}+q_{x}^{*}{{q}_{xxxx}}+q\big[10{{({{q}^{*}})}^{2}}{{q}_{xx}}+q_{xxxx}^{*}\big]\big\} \\&\quad\quad+{{A}_{5}}\big\{-6{{q}^{2}}{{q}^{*}}q_{x}^{*}-q_{x}^{*}{{q}_{xx}}+{{q}_{x}}q_{xx}^{*}+{{q}^{*}}{{q}_{xxx}}+q\big[6{{({{q}^{*}})}^{2}}{{q}_{x}}-q_{xxx}^{*}\big]\big\}\\&\quad\quad-i{{A}_{4}}\big\{3{{q}^{2}}{{({{q}^{*}})}^{2}}-{{q}_{x}}q_{x}^{*} +{{q}^{*}}{{q}_{xx}}+qq_{xx}^{*}\big\}+{{A}_{3}}({{q}^{*}}{{q}_{x}}-qq_{x}^{*})-i{{A}_{2}}q{{q}^{*}},\\
 & {{a}_{1}}=2{{A}_{8}}\big\{30{{q}_{xxx}}{{({{q}^{*}})}^{2}}q_{x}^{*}-20{{({{q}^{*}})}^{2}}{{q}_{x}}{{q}_{xx}}-q_{xx}^{*}{{q}_{xxx}}+10{{q}^{2}}(-3{{({{q}^{*}})}^{3}}{{q}_{x}}+2q_{x}^{*}q_{xx}^{*}\\&\quad\quad+{{q}^{*}}q_{xxx}^{*})+{{q}_{xx}}q_{xxx}^{*}q_{x}^{*}{{q}_{xxxx}} -{{q}_{x}}q_{xxxx}^{*}-{{q}^{*}}(10q_{x}^{2}q_{x}^{*}+{{q}_{xxxxx}})+q\big[10{{q}^{*}}q_{x}^{*}{{q}_{xx}}\\&\quad\quad +10{{q}_{x}}({{(q_{x}^{*})}^{2}}-{{q}^{*}}q_{xx}^{*})-10{{({{q}^{*}})}^{2}}{{q}_{xxx}}+q_{xxxxx}^{*}\big]\big\}-2i{{A}_{7}}\big\{10{{\left| q \right|}^{6}}+5{{({{q}^{*}})}^{2}}q_{x}^{2}\\&\quad\quad+{{q}_{xx}}q_{xx}^{*}+5{{q}^{2}}({{(q_{x}^{*})}^{2}}+2{{q}^{*}}q_{xx}^{*}) -q_{x}^{*}{{q}_{xxx}}-{{q}_{x}}q_{xxx}^{*}+{{q}^{*}}{{q}_{xxxx}}+q(10{{({{q}^{*}})}^{2}}{{q}_{xx}}\\&\quad\quad+q_{xxxx}^{*})\big\}+2{{A}_{6}}\big\{6{{q}^{2}}{{q}^{*}}q_{x}^{*}+q_{x}^{*}{{q}_{xx}}-{{q}_{x}}q_{xx}^{*}-q_{x}^{*}{{q}_{xxx}}+q(-6{{({q_{x}^{*}})}^{2}}{{q}_{x}}+q_{xxx}^{*})\big\}\\&\quad\quad-2i{{A}_{5}}\big\{3{{\left| q \right|}^{4}}-{{q}_{x}}q_{x}^{*}+{{q}^{*}}{{q}_{xx}}+qq_{xx}^{*}\big\}+2{{A}_{4}}\big\{{{q}_{x}}{{q}^{*}}+qq_{x}^{*}\big\}-2i{{A}_{3}}q{{q}^{*}},\\
 & {{a}_{2}}=4i{{A}_{8}}\big\{10{{\left| q \right|}^{6}}+5{{({{q}^{*}})}^{2}}q_{x}^{2}+{{q}_{xx}}q_{xx}^{*}+5{{q}^{2}}({{(q_{x}^{*})}^{2}}+2{{q}^{*}}q_{xx}^{*})-q_{x}^{*}{{q}_{xxx}}-{{q}_{x}}q_{xxx}^{*}\\&\quad \quad+{{q}^{*}}{{q}_{xxxx}} +q(10{{({{q}^{*}})}^{2}}{{q}_{xx}}+q_{xxxx}^{*})\big\}+4{{A}_{7}}\big\{6{{q}^{2}}{{q}^{*}}q_{x}^{*}+q_{x}^{*}{{q}_{xx}}-{{q}_{x}}q_{xx}^{*}\\&\quad\quad-{{q}^{*}}{{q}_{xxx}} +q(-6{{({{q}^{*}})}^{2}}{{q}_{x}}+q_{xxx}^{*})\big\}+4i{{A}_{6}}\big\{3{{\left| q \right|}^{4}}-{{q}_{x}}q_{x}^{*}+{{q}^{*}}{{q}_{xx}}+qq_{xx}^{*}\big\}\\&\quad\quad+4{{A}_{5}}\big\{{{q}_{x}}{{q}^{*}}-qq_{x}^{*}\big\}+4i{{A}_{4}}q{{q}^{*}}+2i{{A}_{2}},\\
 & {{a}_{3}}=-8{{A}_{8}}\big\{6{{q}^{2}}{{q}^{*}}q_{x}^{*}+q_{x}^{*}{{q}_{xx}}-{{q}_{x}}q_{xx}^{*}-{{q}^{*}}{{q}_{xxx}}+q(-6{{({{q}^{*}})}^{2}}{{q}_{x}}+q_{xxx}^{*})\big\}\\
 &\quad\quad+8i{{A}_{7}}\big\{3{{q}^{2}}{{({{q}^{*}})}^{2}}-{{q}_{x}}q_{x}^{*}+q_{xx}{{q}^{*}}+qq_{xx}^{*}\big\}-8{{A}_{6}}\big\{-{{q}_{x}}{{q}^{*}}+qq_{x}^{*}\big\}\\&\quad\quad+8i{{A}_{5}}q{{q}^{*}}+4i{{A}_{3}},\\
 & {{a}_{4}}=-16i{{A}_{8}}\big\{3{{\left| q \right|}^{4}}-{{q}_{x}}q_{x}^{*}+{{q}^{*}}{{q}_{xx}}+qq_{xx}^{*}\big\}+16{{A}_{7}}\big\{{{{q}_{x}}{{q}^{*}}-qq_{x}^{*}}\big\}-16i{{A}_{6}}q{{q}^{*}}-8i{{A}_{4}},\\
 & {{a}_{5}}=32{{A}_{8}}\big\{-{q}_{x}{{q}^{*}}+qq_{x}^{*}\big\}-32i{{A}_{7}}q{{q}^{*}}-16i{{A}_{5}},\\
 & {{a}_{6}}=64i{{A}_{8}}q{{q}^{*}}+32i{{A}_{6}},\quad {{a}_{7}}=64i{{A}_{7}},\quad {{a}_{8}}=-128i{{A}_{8}},\end{aligned}\]\[\begin{aligned}&{{b}_{0}}={{A}_{8}}\big\{140{{\left| q \right|}^{6}}{{q}_{x}}+70{{({{q}^{*}})}^{2}}q_{x}^{3}+70q_{xx}^{2}q_{x}^{*}+112{{q}_{x}}{{q}_{xx}}q_{xx}^{*}+98{{q}_{x}}{{q}_{xxx}}q_{x}^{*} \\
 &\quad\quad+70{{q}^{2}}({{q}_{x}}{{(q_{x}^{*})}^{2}}+2{{q}^{*}}q_{xx}^{*})+{{q}^{*}}(2{{q}_{xx}}q_{x}^{*}+{{q}^{*}}{{q}_{xxx}})+28q_{x}^{2}q_{xxx}^{*} \\
 &\quad\quad +14{{q}^{*}}(5{{q}_{xx}}{{q}_{xxx}}+3{{q}_{x}}{{q}_{xxxx}})+14q(20{{({{q}^{*}})}^{2}}{{q}_{x}}{{q}_{xx}}+3q_{xx}^{*}{{q}_{xxx}}+{{q}_{xx}}q_{xxx}^{*} \\
 &\quad\quad +2q_{x}^{*}{{q}_{xxxx}}+{{q}_{x}}q_{xxxx}^{*}+{{q}^{*}}(20q_{x}^{2}q_{x}^{*}+{{q}_{xxxxx}}))+{{q}_{xxxxxxx}}\big\}+i{{A}_{7}}\big\{20{{q}^{4}}{{({{q}^{*}})}^{3}} \\
 &\quad\quad +20{{q}^{*}}q_{xx}^{2}+20q_{x}^{2}q_{xx}^{*}+10{{q}^{3}}(5q_{x}^{*}{{q}_{xx}}+3{{q}_{xxx}}{{q}^{*}})+2q(35{{({{q}^{*}})}^{2}}q_{x}^{2}+11{{q}_{xx}}q_{xx}^{*} \\
 &\quad\quad+9{{q}^{*}}{{q}_{xxx}}+4{{q}_{x}}q_{xxx}^{*}+6{{q}_{xxxx}}{{q}^{*}})+2{{q}^{2}}(30{{q}^{*}}{{q}_{x}}q_{x}^{*}+25{{({{q}^{*}})}^{2}}{{q}_{xx}}+q_{xxxx}^{*}) \\
 &\quad\quad+{{q}_{xxxxxx}}\big\}+{{A}_{6}}\big\{30{{\left| q \right|}^{4}}{{q}_{x}}+10q_{x}^{2}q_{x}^{*}+20{{q}^{*}}{{q}_{x}}{{q}_{xx}}+10q(q_{x}^{*}{{q}_{xx}}+{{q}_{x}}q_{xx}^{*} \\
 &\quad\quad+{{q}^{*}}{{q}_{xxx}})+{{q}_{xxxxx}}\big\}+i{{A}_{5}}\big\{6q{{\left| q \right|}^{4}}+6{{q}^{*}}q_{x}^{2}+4q({{q}_{x}}q_{x}^{*}+2{{q}^{*}}{{q}_{xx}})+2{{q}^{2}}q_{xx}^{*} \\
 &\quad\quad+{{q}_{xxxx}}\big\}+{{A}_{4}}\big\{6q{{q}^{*}}{{q}_{x}}+{{q}_{xxx}}\big\}+i{{A}_{3}}\big\{2{{q}^{2}}{{q}^{*}}+{{q}_{xx}}\big\}+{{A}_{2}}{{q}_{x}},\\
 &{{b}_{1}}=-2i{{A}_{8}}\big\{20q{{\left| q \right|}^{6}}+20{{q}^{*}}q_{xx}^{2}+20q_{x}^{2}q_{xx}^{*}+10{{q}^{3}}({{(q_{x}^{*})}^{2}}+2{{q}^{*}}q_{xx}^{*})+10{{q}_{x}}(5q_{x}^{*}{{q}_{xx}} \\
 &\quad\quad+3q_{x}^{*}{{q}_{xxx}})+2q(35{{({{q}^{*}}{{q}_{x}})}^{2}}+11{{q}_{xx}}q_{xx}^{*}+9q_{x}^{*}{{q}_{xxx}}+4{{q}_{x}}q_{xxx}^{*}+6{{q}^{*}}{{q}_{xxxx}}) \\
 &\quad\quad+2{{q}^{2}}(30{{q}^{*}}{{q}_{x}}q_{x}^{*}+25{{({{q}^{*}})}^{2}}{{q}_{xx}}+q_{xxxx}^{*})+{{q}_{xxxxxx}}\big\}+2{{A}_{7}}\big\{30{{\left| q \right|}^{4}}{{q}_{x}}+10q_{x}^{2}q_{x}^{*} \\
 &\quad\quad+20{{q}^{*}}{{q}_{x}}{{q}_{xx}}+10q(q_{x}^{*}{{q}_{xx}}+{{q}_{x}}q_{xx}^{*}+{{q}^{*}}{{q}_{xxx}})+{{q}_{xxxxx}}\big\}-2i{{A}_{6}}\big\{6q{{\left| q \right|}^{4}} \\
 &\quad\quad+6{{q}^{*}}q_{x}^{2}+4q({{q}_{x}}q_{x}^{*}+2{{q}^{*}}{{q}_{xx}})+2{{q}^{2}}q_{xx}^{*}+{{q}_{xxxx}}\big\}+2{{A}_{5}}\big\{6q{{q}^{*}}{{q}_{x}}+{{q}_{xxx}}\big\} \\
 &\quad\quad-2i{{A}_{4}}\big\{2{{q}^{2}}{{q}^{*}}+{{q}_{xx}}\big\}+2{{A}_{3}}{{q}_{x}}-2i{{A}_{2}}q,\\
 &{{b}_{2}}=-4{{A}_{8}}\big\{30{{\left| q \right|}^{4}}{{q}_{x}}+10q_{x}^{2}q_{x}^{*}+20{{q}^{*}}{{q}_{x}}{{q}_{xx}}+10q(q_{x}^{*}{{q}_{xx}}+{{q}_{x}}q_{xx}^{*}+{{q}^{*}}{{q}_{xxx}}) \\
 &\quad\quad+{{q}_{xxxxx}}\big\}-4i{{A}_{7}}\big\{6q{{\left| q \right|}^{4}}+6{{q}^{*}}q_{x}^{2}+4q({{q}_{x}}q_{x}^{*}+2{{q}^{*}}{{q}_{xx}})+2{{q}^{2}}q_{xx}^{*}+{{q}_{xxxx}}\big\} \\
 &\quad\quad-4{{A}_{6}}\big\{6{{\left| q \right|}^{2}}{{q}_{x}}+{{q}_{xxx}}\big\}-4i{{A}_{5}}\big\{2q{{\left| q \right|}^{2}}+{{q}_{xx}}\big\}-4{{A}_{4}}{{q}_{x}}-4i{{A}_{3}}q, \\
 &{{b}_{3}}=8i{{A}_{8}}\big\{6q{{\left| q \right|}^{4}}+6{{q}^{*}}q_{x}^{2}+4q({{q}_{x}}q_{x}^{*}+2{{q}^{*}}{{q}_{xx}})+2{{q}^{2}}q_{xx}^{*}+{{q}_{xxxx}}\big\} \\
 &\quad\quad-8{{A}_{7}}\big\{6{{\left| q \right|}^{2}}{{q}_{x}}+{{q}_{xxx}}\big\}+8i{{A}_{6}}\big\{2q{{\left| q \right|}^{2}}+{{q}_{xx}}\big\}-8{{A}_{5}}{{q}_{x}}+8i{{A}_{4}}q, \\
 &{{b}_{4}}=16{{A}_{8}}\big\{6{{\left| q \right|}^{2}}{{q}_{x}}+{{q}_{xxx}}\big\}+16i{{A}_{7}}\big\{2q{{\left| q \right|}^{2}}+{{q}_{xx}}\big\}+16{{A}_{6}}{{q}_{x}}+16i{{A}_{5}}q, \\
 & {{b}_{5}}=-32i{{A}_{8}}\big\{2q{{\left| q \right|}^{2}}+{{q}_{xx}}\big\}+32{{A}_{7}}{{q}_{x}}-32i{{A}_{6}}q,\quad {{b}_{6}}=-64{{A}_{8}}{{q}_{x}}-64i{{A}_{7}}q, \\
 & {{b}_{7}}=128i{{A}_{8}}q,\quad {{b}_{8}}=0,\quad {{c}_{j}}=b_{j}^{*}.
\end{aligned}\]

Let us now rewrite the Lax pair (2) in a more convenient form
\begin{subequations}
\begin{align}
 & {{\Psi }_{x}}=i(\varsigma \sigma +Q)\Psi , \\
 & {{\Psi }_{t}}=\big[i\big(2{{A}_{2}}{{\varsigma}^{2}}+4{{A}_{3}}{{\varsigma}^{3}}-8{{A}_{4}}{{\varsigma}^{4}}-16{{A}_{5}}{{\varsigma}^{5}}+32{{A}_{6}}{{\varsigma }^{6}}+64{{A}_{7}}{{\varsigma}^{7}}-128{{A}_{8}}{{\varsigma}^{8}}\big)\sigma +{{Q}_{1}}\big]\Psi ,
\end{align}
\end{subequations}
where
\[
\sigma =\left( \begin{matrix}
   1 & 0  \\
   0 & -1  \\
\end{matrix} \right),\quad
Q=\left( \begin{matrix}
   0 & {q}^{*}  \\
   q & 0  \\
\end{matrix} \right),\]\[{{Q}_{1}}=\big({{a}_{0}}+{{a}_{1}}\varsigma+{{\hat{a}}_{2}}{{\varsigma}^{2}}+{{\hat{a}}_{3}}{{\varsigma}^{3}}+{{\hat{a}}_{4}}{{\varsigma }^{4}}+{{\hat{a}}_{5}}{{\varsigma}^{5}}+{{\hat{a}}_{6}}{{\varsigma}^{6}}\big)\sigma +\sum\limits_{j=0}^{8}{{{\varsigma}^{j}}\left( \begin{matrix}
   0 & {{b}_{j}}  \\
   {{c}_{j}} & 0  \\
\end{matrix} \right)},
\]
and ${\hat{a}}_{l}$ mean $q(x,t)$ and its derivative terms appeared in ${a}_{l}(2\leq l\leq6)$.

\section{Riemann-Hilbert problem}
In this section, we focus on putting forward a Riemann-Hilbert problem for the eighth-order NLS equation (1). Now we assume that the potential function $q(x,t)$ in the Lax pair (3) decays to zero sufficiently fast as $x\rightarrow\pm\infty$. It can be known from (3) that when $x\rightarrow\pm\infty$,
\[\Psi \propto{{\text{e}}^{i\varsigma \sigma x+i(2{{A}_{2}}{{\varsigma}^{2}}+4{{A}_{3}}{{\varsigma}^{3}}-8{{A}_{4}}{{\varsigma}^{4}}-16{{A }_{5}}{{\varsigma}^{5}}+32{{A}_{6}}{{\varsigma}^{6}}+64{{A}_{7}}{{\varsigma}^{7}}-128{{A}_{8}}{{\zeta }^{8}})\sigma t}},\]
which motivates us to introduce the variable transformation
\[
\Psi=\mu{{\text{e}}^{i\varsigma \sigma x+i(2{{A}_{2}}{{\varsigma}^{2}}+4{{A}_{3}}{{\varsigma}^{3}}-8{{A}_{4}}{{\varsigma}^{4}}-16{{A }_{5}}{{\varsigma}^{5}}+32{{A}_{6}}{{\varsigma}^{6}}+64{{A}_{7}}{{\varsigma}^{7}}-128{{A}_{8}}{{\varsigma}^{8}})\sigma t}}.
\]
Upon this transformation, the Lax pair (3) can be changed into the desired form
\begin{subequations}
\begin{align}
 & {{\mu}_{x}}=i\varsigma[\sigma ,\mu]+U_{1}\mu, \\
 & {{\mu}_{t}}=i\big(2{{A}_{2}}{{\varsigma}^{2}}+4{{A}_{3}}{{\varsigma}^{3}}-8{{A}_{4}}{{\varsigma}^{4}}-16{{A }_{5}}{{\varsigma}^{5}}+32{{A}_{6}}{{\varsigma}^{6}}+64{{A}_{7}}{{\varsigma}^{7}}-128{{A}_{8}}{{\varsigma}^{8}}\big)[\sigma ,\mu]+{{Q}_{1}}\mu,
\end{align}
\end{subequations}
where $[\cdot,\cdot]$ is the matrix commutator and
$U_{1}=iQ.$
From (4), we find that $\text{tr}({{U}_{1}})=\text{tr}({{Q}_{1}})=0$.

In the direct scattering process, we will concentrate on the spectral problem (4a), and
the $t$-dependence will be suppressed.
We first introduce two matrix Jost solutions $\mu_{\pm}$ of (4a) expressed as a collection of columns
\begin{equation}
{{\mu}_{-}}=({{[{{\mu}_{-}}]}_{1}},{{[{{\mu}_{-}}]}_{2}}),\quad {{\mu}_{+}}=({{[{{\mu}_{+}}]}_{1}},{{[{{\mu}_{+}}]}_{2}})
\end{equation}
meeting the asymptotic conditions
\begin{subequations}
\begin{align}
&{{\mu}_{-}}\to \mathbb{I},\quad x\to -\infty ,\\
&{{\mu}_{+}}\to \mathbb{I},\quad x\to +\infty .
\end{align}
\end{subequations}
Here the subscripts of $\mu$ indicated refer to which end of the $x$-axis the boundary conditions are required for, and $\mathbb{I}$ stands for the identity matrix of size 2. Actually, the solutions ${{\mu}_{\pm }}$ are uniquely determined by the integral equations of Volterra type
\begin{subequations}
\begin{align}
&{{\mu}_{-}}=\mathbb{I}+\int_{-\infty }^{x}{{{\text{e}}^{i\varsigma \sigma (x-y)}}{{U}_{1}}(y){{\mu}_{-}}(y,\varsigma){{\text{e}}^{i\varsigma \sigma (y-x )}}\text{d}y},\\
&{{\mu}_{+}}=\mathbb{I}-\int_{x}^{+\infty }{{{\text{e}}^{i\varsigma \sigma (x-y)}}{{U}_{1}}(y){{\mu}_{+}}(y,\varsigma){{\text{e}}^{i\varsigma \sigma (y-x )}}\text{d}y}.
\end{align}
\end{subequations}
After direct analysis on Equations (7) we can see that ${{[{{\mu}_{-}}]}_{1}},{{[{{\mu}_{+}}]}_{2}}$ are analytic for $\varsigma \in {\mathbb{C}^{-}}$ and continuous for $\varsigma \in {\mathbb{C}^{-}}\cup \mathbb{R}$, while ${{[{{\mu}_{+}}]}_{1}},{{[{{\mu}_{-}}]}_{2}}$ are analytic for $\varsigma \in {\mathbb{C}^{+}}$ and continuous for $\varsigma \in {\mathbb{C}^{+}}\cup \mathbb{R}$, where ${\mathbb{C}^{-}}$ and ${\mathbb{C}^{+}}$ are respectively the lower and upper half $\varsigma$-plane:
\[{{\mathbb{C}}^{-}}=\left\{ \varsigma \in \mathbb{C}|\operatorname{Im}(\varsigma )<0 \right\},\quad {{\mathbb{C}}^{+}}=\left\{ \varsigma \in \mathbb{C}|\operatorname{Im}(\varsigma )>0 \right\}.\]

Next we set out to study the properties of $\mu_{\pm}$. In fact,
it can be shown from $\text{tr}({{U}_{1}})=0$ that the determinants of ${{\mu}_{\pm }}$ are independent of the variable $x$. Evaluating $\det {{\mu}_{-}}$ at $x=-\infty$ and $\det {{\mu}_{+}}$ at $x=+\infty$, we get $\det {{\mu}_{\pm }}=1$ for $\varsigma \in \mathbb{R}.$
In addition, ${{\mu}_{-}}E$ and ${{\mu}_{+}}E$ are both fundamental solutions of (3a),
where $E={{\text{e}}^{i\varsigma \sigma x}}$,
they are linearly dependent
\begin{equation}
{{\mu}_{-}}E={{\mu}_{+}}ES(\varsigma),\quad \varsigma \in \mathbb{R}.
\end{equation}
Here $S(\varsigma)={{({{s}_{kj}})}_{2\times 2}}$ is called the scattering matrix and
$\det S(\varsigma)=1.$
Furthermore, we find from the properties of $\mu_{\pm}$ that ${{s}_{11}}$ allows analytic extension to ${\mathbb{C}^{-}}$ and ${{s}_{22}}$ analytically extends to ${\mathbb{C}^{+}}$.

A Riemann-Hilbert problem desired is closely associated with two matrix functions: one is analytic in ${\mathbb{C}^{+}}$ and the other is analytic in ${\mathbb{C}^{-}}$. In consideration of the analytic properties of $\mu_{\pm}$, we set
\begin{equation}
{{P}_{1}}(x,\varsigma)=({{[{{\mu}_{+}}]}_{1}},{{[{{\mu}_{-}}]}_{2}})(x,\varsigma),
\end{equation}
defining in ${\mathbb{C}^{+}}$, be an analytic function of $\varsigma$. And then, ${{P}_{1}}$ can be expanded into the asymptotic series at large-$\varsigma$
\begin{equation}
{{P}_{1}}=P_{1}^{(0)}+\frac{P_{1}^{(1)}}{\varsigma}+\frac{P_{1}^{(2)}}{{{\varsigma}^{2}}}+O\bigg( \frac{1}{{{\varsigma}^{3}}} \bigg),\quad \varsigma \to \infty .
\end{equation}
Inserting expansion (10) into the spectral problem (4a) and equating terms with same powers of $\varsigma$, we obtain
\begin{equation}
i\big[\sigma ,P_{1}^{(1)}\big]+{{U}_{1}}P_{1}^{(0)}=P_{1x}^{(0)},\quad
i\big[\sigma ,P_{1}^{(0)}\big]=0, \nonumber
\end{equation}
which yields $P_{1}^{(0)}=\mathbb{I}$, namely
${{P}_{1}}\to \mathbb{I}$ as $ \varsigma \in {\mathbb{C}^{+}}\to \infty .$

For establishing a Riemann-Hilbert problem, the analytic counterpart of $P_{1}$ in ${\mathbb{C}^{-}}$ is still needed to be given.
Noting that the adjoint scattering equation of (4a) reads as
\begin{equation}
{{H}_{x}}=i\varsigma[\sigma ,H]-H{{U}_{1}},
\end{equation}
and the inverse matrices of ${{\mu}_{\pm }}$ meet this adjoint equation.
Then we express
the inverse matrices of ${{\mu}_{\pm }}$ as a collection of rows
\begin{equation}
\mu_{\pm}^{-1}=\left( \begin{matrix}
   {[\mu_{\pm}^{-1}]^{1}}  \\
   {[\mu_{\pm}^{-1}]^{2}}  \\
\end{matrix} \right),
\end{equation}
which obey the boundary conditions $\mu_{\pm}^{-1}\rightarrow\mathbb{I}$ as $x\rightarrow\pm\infty$.
It is easy to know from (8) that
\begin{equation}
{{E}^{-1}}\mu_{-}^{-1}=R(\varsigma){{E}^{-1}}\mu_{+}^{-1},
\end{equation}
where $R(\varsigma)={{({{r}_{kj}})}_{2\times 2}}={{S}^{-1}}(\varsigma)$.
Thus, the matrix function ${{P}_{2}}$ which is analytic for $\varsigma \in {\mathbb{C}^{-}}$ is constructed as
\begin{equation}
{{P}_{2}}(x,\varsigma)=\left( \begin{matrix}
   {[\mu_{+}^{-1}]^{1}}  \\
   {[\mu_{-}^{-1}]^{2}}  \\
\end{matrix} \right)(x,\varsigma).
\end{equation}
Analogous to ${{P}_{1}}$, the very large-$\varsigma$ asymptotic behavior of ${{P}_{2}}$ turns out to be
$
{{P}_{2}}\to \mathbb{I}$ as $\varsigma \in {\mathbb{C}^{-}}\to \infty .
$

Carrying (5) into Equation (8) gives rise to
\begin{equation*}
({{[{{\mu}_{-}}]}_{1}},{{[{{\mu}_{-}}]}_{2}})=({{[{{\mu}_{+}}]}_{1}},{{[{{\mu}_{+}}]}_{2}})\left( \begin{matrix}
   {{s}_{11}} & {{s}_{12}}{{\text{e}}^{2i\varsigma x}}  \\
   {{s}_{21}}{{\text{e}}^{-2i\varsigma x}} & {{s}_{22}}  \\
\end{matrix} \right),
\end{equation*}
from which we have
\begin{equation*}
{{[{{\mu}_{-}}]}_{2}}={{s}_{12}}{{\text{e}}^{2i\varsigma x}}{{[{{\mu}_{+}}]}_{1}}+{{s}_{22}}{{[{{\mu}_{+}}]}_{2}}.
\end{equation*}
Hence, ${{P}_{1}}$ is of the form
\begin{equation*}
{{P}_{1}}=({{[{{\mu}_{+}}]}_{1}},{{[{{\mu}_{-}}]}_{2}})=({{[{{\mu}_{+}}]}_{1}},{{[{{\mu}_{+}}]}_{2}})\left( \begin{matrix}
   1 & {{s}_{12}}{{\text{e}}^{2i\varsigma x}}  \\
   0 & {{s}_{22}}  \\
\end{matrix} \right).
\end{equation*}

On the other hand, via substituting (12) into Equation (13), we get
\begin{equation*}
\left( \begin{matrix}
   {[\mu_{-}^{-1}]^{1}}  \\
   {[\mu_{-}^{-1}]^{2}}  \\
\end{matrix} \right)=\left( \begin{matrix}
   {{r}_{11}} & {{r}_{12}}{{\text{e}}^{2i\varsigma x}}  \\
   {{r}_{21}}{{\text{e}}^{-2i\varsigma x}} & {{r}_{22}}  \\
\end{matrix} \right)\left( \begin{matrix}
   {[\mu_{+}^{-1}]^{1}}  \\
   {[\mu_{+}^{-1}]^{2}}  \\
\end{matrix} \right),
\end{equation*}
from which we can express ${[\mu_{-}^{-1}]^{2}}$ as
\begin{equation*}
{[\mu_{-}^{-1}]^{2}}={{r}_{21}}{{\text{e}}^{-2i\varsigma x}}{[\mu_{+}^{-1}]^{1}}+{{r}_{22}}{[\mu_{+}^{-1}]^{2}}.
\end{equation*}
As a consequence, ${{P}_{2}}$ is written as
\begin{equation*}
{{P}_{2}}=\left( \begin{matrix}
   {[\mu_{+}^{-1}]^{1}}  \\
   {[\mu_{-}^{-1}]^{2}}  \\
\end{matrix} \right)=\left( \begin{matrix}
   1 & 0  \\
   {{r}_{21}}{{\text{e}}^{-2i\varsigma x}} & {{r}_{22}}  \\
\end{matrix} \right)\left( \begin{matrix}
   {[\mu_{+}^{-1}]^{1}}  \\
   {[\mu_{+}^{-1}]^{2}}  \\
\end{matrix} \right).
\end{equation*}

With two matrix functions ${{P}_{1}}$ and ${{P}_{2}}$ which are analytic in ${\mathbb{C}^{+}}$ and ${\mathbb{C}^{-}}$ respectively in hand,
we are in a position to deduce a Riemann-Hilbert problem for the eighth-order NLS equation (1).
After denoting that the limit of ${{P}_{1}}$ is ${{P}^{+}}$ as $\varsigma \in {\mathbb{C}^{+}}\rightarrow\mathbb{R}$ and
the limit of ${{P}_{2}}$ is ${{P}^{-}}$ as $\varsigma \in {\mathbb{C}^{-}}\rightarrow\mathbb{R}$, a Riemann-Hilbert problem can be given below
\begin{equation}
{{P}^{-}}(x,\varsigma){{P}^{+}}(x,\varsigma)=\left( \begin{matrix}
   1 & {{s}_{12}}{{\text{e}}^{2i\varsigma x}}  \\
   {{r}_{21}}{{\text{e}}^{-2i\varsigma x}} & 1  \\
\end{matrix} \right),
\end{equation}
with its canonical normalization conditions as
\begin{eqnarray*}
  &{{P}_{1}}(x,\varsigma)\to \mathbb{I},\quad \varsigma \in {\mathbb{C}^{+}}\to \infty , \\
  &{{P}_{2}}(x,\varsigma)\to \mathbb{I},\quad \varsigma \in {\mathbb{C}^{-}}\to \infty ,
\end{eqnarray*}
and ${{r}_{21}}{{s}_{12}}+{{r}_{22}}{{s}_{22}}=1$.

\section{$N$-soliton solution}
Having described a Riemann-Hilbert problem for Equation (1), we now turn to construct its multi-soliton solutions. To achieve the goal, we first need to solve the Riemann-Hilbert problem (15) under the assumption of irregularity,
which signifies that both $\det {{P}_{1}}$ and $\det {{P}_{2}}$ possess some zeros in the analytic domains of their own.
From the definitions of ${{P}_{1}}$ and ${{P}_{2}}$, we have
\begin{align}
& \det {{P}_{1}}(\varsigma)={{s}_{22}}(\varsigma),\quad \varsigma \in {\mathbb{C}^{+}},  \nonumber \\
& \det {{P}_{2}}(\varsigma)={{r}_{22}}(\varsigma),\quad \varsigma \in {\mathbb{C}^{-}},  \nonumber
\end{align}
which means that
$\det {{P}_{1}}$ and $\det {{P}_{2}}$ have the same zeros as ${s}_{22}$ and ${r}_{22}$ respectively, and ${{r}_{22}}={{({{S}^{-1}})}_{22}}={{s}_{11}}$.

With above analysis, it is now necessary to reveal the characteristic feature of zeros.
It can be noticed that the potential matrix $Q$ has the symmetry property
$Q^{\dagger }=Q,$
upon which we deduce
\begin{equation}
\mu_{\pm }^{\dagger }({{\varsigma}^{*}})=\mu_{\pm }^{-1}(\varsigma).
\end{equation}
Here the subscript $\dagger$ stands for the Hermitian of a matrix.
In order to facilitate discussion, we introduce two special matrices ${{H}_{1}}=\text{diag}(1,0)$ and ${{H}_{2}}=\text{diag}(0,1),$ and
express (9) and (14) in terms of
\begin{subequations}
\begin{align}
&{{P}_{1}}={{\mu}_{+}}{{H}_{1}}+{{\mu}_{-}}{{H}_{2}},\\
&{{P}_{2}}={{H}_{1}}\mu_{+}^{-1}+{{H}_{2}}\mu_{-}^{-1}.
\end{align}
\end{subequations}
A direct computation of the Hermitian of expression (17a), using the relation (16), generates that
\begin{equation}
P_{1}^{\dagger }({{\varsigma}^{*}})={{P}_{2}}(\varsigma),\quad \varsigma \in {\mathbb{C}^{-}},
\end{equation}
and the involution property of scattering matrix
$
{{S}^{\dagger }}({{\varsigma}^{*}})={{S}^{-1}}(\varsigma),
$
which leads to
\begin{equation}
s_{22}^{*}({{\varsigma}^{*}})={{r}_{22}}(\varsigma),\quad \varsigma \in {\mathbb{C}^{-}}.
\end{equation}
This equality implies that each zero $\pm {{\varsigma}_{k}}$ of ${{s}_{22}}$ results in each zero $\pm \varsigma_{k}^{*}$ of ${{r}_{22}}$ correspondingly.
Therefore, our assumption is that $\det {{P}_{1}}$ has simple zeros $\{{{\varsigma }_{j}}\in {{\mathbb{C}}^{+}},1\le j\le N\}$
and $\det {{P}_{2}}$ has simple zeros $\{{{\hat{\varsigma }}_{j}}\in {{\mathbb{C}}^{-}},1\le j\le N\}$, where
${{\hat{\varsigma}}_{l}}=\varsigma_{l}^{*},$ $ 1\le l\le N.$
The full set of the discrete scattering data is composed of these zeros and the nonzero column vectors ${{\upsilon}_{j}}$ and row vectors ${{\hat{\upsilon}}_{j}}$, which satisfy the following equations
\begin{subequations}
\begin{align}
&{{P}_{1}}({{\varsigma}_{j}}){{\upsilon}_{j}}=0,\\
&{{\hat{\upsilon}}_{j}}{{P}_{2}}({{\hat{\varsigma}}_{j}})=0.
\end{align}
\end{subequations}

Taking the Hermitian of Equation (20a) and using (18) as well as comparing with Equation (20b), we find that the eigenvectors fulfill the relation
\begin{equation}
{{\hat{\upsilon}}_{j}}=\upsilon_{j}^{\dagger },\quad 1\le j\le N.
\end{equation}
Differentiating Equation (20a) about $x$ and $t$ and taking advantage of the Lax pair (4), we arrive at
\begin{equation*}
\begin{aligned}
 & {{P}_{1}}({{\varsigma}_{j}})\left( \frac{\partial {{\upsilon }_{j}}}{\partial x}-i{{\varsigma}_{j}}\sigma {{\upsilon}_{j}} \right)=0, \\
 & {{P}_{1}}({{\varsigma}_{j}})\left( \frac{\partial {{\upsilon }_{j}}}{\partial t}-i\big(2{{A}_{2}}{\varsigma _{j}^{2}}+4{{A}_{3}}{\varsigma _{j}^{3}}-8{{A}_{4}}{\varsigma _{j}^{4}}-16{{A}_{5}}{\varsigma _{j}^{5}}+32{{A}_{6}}{\varsigma _{j}^{6}}+64{{A}_{7}}{\varsigma _{j}^{7}}-128{{A}_{8}}{\varsigma _{j}^{8}}\big)\sigma{{\upsilon}_{j}} \right)=0,
\end{aligned}
\end{equation*}
which yields
\begin{equation*}
{{\upsilon}_{j}}\text{=}{{\text{e}}^{(i{{\varsigma}_{j}}x+i(2{{A}_{2}}{\varsigma _{j}^{2}}+4{{A}_{3}}{\varsigma _{j}^{3}}-8{{A}_{4}}{\varsigma _{j}^{4}}-16{{A}_{5}}{\varsigma _{j}^{5}}+32{{A}_{6}}{\varsigma _{j}^{6}}+64{{A}_{7}}{\varsigma _{j}^{7}}-128{{A}_{8}}{\varsigma _{j}^{8}})t)\sigma }}{{\upsilon}_{j,0}},\quad 1\le j\le N.
\end{equation*}
Here ${{\upsilon}_{j,0}},1\le j\le N,$ are complex constant vectors. Making use of the relation (21), we have
\begin{equation*}
{{\hat{\upsilon }}_{j}}=\upsilon _{j,0}^{\dagger }{{\text{e}}^{(-i\varsigma _{j}^{*}x-i(2{{A}_{2}}\varsigma {{_{j}^{*}}^{2}}+4{{A}_{3}}\varsigma {{_{j}^{*}}^{3}}-8{{A}_{4}}\varsigma {{_{j}^{*}}^{4}}-16{{A}_{5}}\varsigma {{_{j}^{*}}^{5}}+32{{A}_{6}}\varsigma {{_{j}^{*}}^{6}}+64{{A}_{7}}\varsigma {{_{j}^{*}}^{7}}-128{{A}_{8}}\varsigma {{_{j}^{*}}^{8}})t)\sigma }},\quad 1\le j\le N.
\end{equation*}

However, in order to derive soliton solutions of the eighth-order NLS equation (1), we investigate the Riemann-Hilbert problem (15) corresponding to the reflectionless case, i.e., ${{s}_{12}}=0$. Introducing a $N\times N$ matrix $M$ defined as
\begin{equation*}
M=({{M}_{kj}})_{N\times N}=\left(\frac{{\hat{\upsilon}_{k}}{{{{\upsilon}}}_{j}}}{{{\varsigma}_{j}}-{{{\hat{\varsigma}}}_{k}}}\right)_{N\times N},\quad 1\le k,j\le N,
\end{equation*}
thus the solution to the problem (15) can be determined by
\begin{subequations}
\begin{align}
& {{P}_{1}}(\varsigma)=\mathbb{I}-\sum\limits_{k=1}^{N}{\sum\limits_{j=1}^{N}{\frac{{{\upsilon}_{k}}{{{\hat{\upsilon}}}_{j}}{{\big({{M}^{-1}}\big)}_{kj}}}{\varsigma -{{{\hat{\varsigma}}}_{j}}}}}, \\
& {{P}_{2}}(\varsigma)=\mathbb{I}+\sum\limits_{k=1}^{N}{\sum\limits_{j=1}^{N}{\frac{{{\upsilon}_{k}}{{{\hat{\upsilon}}}_{j}}{{\big({{M}^{-1}}\big)}_{kj}}}{\varsigma-{{\varsigma }_{k}}}}},
\end{align}
\end{subequations}
where ${{\big({{M}^{-1}}\big)}_{kj}}$ denotes the $(k,j)$-entry of ${{M}^{-1}}$. From expression (22a), it can be seen that
\begin{equation*}
P_{1}^{(1)}=-\sum\limits_{k=1}^{N}{\sum\limits_{j=1}^{N}{{{\upsilon}_{k}}{{{\hat{\upsilon}}}_{j}}{{\big({{M}^{-1}}\big)}_{kj}}}}.
\end{equation*}

In what follows, we shall
retrieve the potential function $q(x,t)$ based on the scattering data. Expanding ${{P}_{1}}(\varsigma)$ at large-$\varsigma$ as
\begin{equation*}
{{P}_{1}}(\varsigma)=\mathbb{I}+\frac{P_{1}^{(1)}}{\varsigma}+\frac{P_{1}^{(2)}}{{{\varsigma}^{2}}}+O\bigg( \frac{1}{{{\varsigma}^{3}}} \bigg),\quad \varsigma \to \infty ,
\end{equation*}
and carrying this expansion into (4a) give rise to
\begin{equation*}
Q=-\big[\sigma ,P_{1}^{(1)}\big].
\end{equation*}
Consequently, the potential function is reconstructed as
\begin{equation*}
q(x,t)=2{{\big(P_{1}^{(1)}\big)}_{21}},
\end{equation*}
with ${{\big(P_{1}^{(1)}\big)}_{21}}$ being the (2,1)-entry of $P_{1}^{(1)}$.

To conclude, setting the nonzero vectors ${{\upsilon}_{k,0}}={({{\alpha }_{k}},{{\beta }_{k}})^\textrm{T}}$ and $\theta_{k}=i{{\varsigma }_{k}}x+i\big(2{{A}_{2}}\varsigma _{k}^{2}+4{{A}_{3}}\varsigma _{k}^{3}-8{{A}_{4}}\varsigma _{k}^{4}-16{{A}_{5}}\varsigma _{k}^{5}+32{{A}_{6}}\varsigma _{k}^{6}+64{{A}_{7}}\varsigma _{k}^{7}-128{{A}_{8}}\varsigma _{k}^{8}\big)t,\operatorname{Im}({{\varsigma }_{k}})>0$,
the general $N$-soliton solution for the eighth-order NLS equation (1) is written as
\begin{equation}
q(x,t)=-2\sum\limits_{k=1}^{N}{\sum\limits_{j=1}^{N}{\alpha _{j}^{*}{{\beta }_{k}}{{\text{e}}^{-{{\theta }_{k}}+\theta _{j}^{*}}}{{\big({{M}^{-1}}\big)}_{kj}}}},
\end{equation}
where
\begin{equation*}
{{M}_{kj}}=\frac{\alpha _{k}^{*}{{\alpha }_{j}}{{\text{e}}^{\theta _{k}^{*}+{{\theta }_{j}}}}+\beta _{k}^{*}{{\beta }_{j}}{{\text{e}}^{-\theta _{k}^{*}-{{\theta }_{j}}}}}{{{\varsigma}_{j}}-\varsigma_{k}^{*}},\quad 1\le k,j\le N.
\end{equation*}

The bright one- and two-soliton solutions will be our main concern in the rest of this section. For the simplest case of $N=1$, the bright one-soliton solution can be readily derived as
\begin{equation}
q(x,t)=-2\alpha _{1}^{*}{{\beta }_{1}}{{\text{e}}^{-{{\theta }_{1}}+\theta _{1}^{*}}}\frac{{{\varsigma}_{1}}-\varsigma_{1}^{*}}{{{\left| {{\alpha }_{1}} \right|}^{2}}{{\text{e}}^{\theta _{1}^{*}+{{\theta }_{1}}}}+{{\left| {{\beta }_{1}} \right|}^{2}}{{\text{e}}^{-\theta _{1}^{*}-{{\theta }_{1}}}}},
\end{equation}
where
$\theta_{1}=i{{\varsigma }_{1}}x+i\big(2{{A}_{2}}\varsigma _{1}^{2}+4{{A}_{3}}\varsigma _{1}^{3}-8{{A}_{4}}\varsigma _{1}^{4}-16{{A}_{5}}\varsigma _{1}^{5}+32{{A}_{6}}\varsigma _{1}^{6}+64{{A}_{7}}\varsigma _{1}^{7}-128{{A}_{8}}\varsigma _{1}^{8}\big)t$.
Furthermore, fixing ${{\beta }_{1}}=1$ and setting ${{\varsigma}_{1}}={{\tilde{a}}_{1}}+i{{\tilde{b}}_{1}}$ as well as ${{\left| {{\alpha }_{1}} \right|}^{2}}={{\text{e}}^{2{{\xi }_{1}}}}$, the solution (24) is then turned into the following form
\begin{equation}
q(x,t)=-2i\alpha _{1}^{*}{{\tilde{b}}_{1}}{{\text{e}}^{-{{\xi }_{1}}}}{{\text{e}}^{\theta _{1}^{*}-{{\theta }_{1}}}}\text{sech}(\theta _{1}^{*}+{{\theta }_{1}}+{{\xi }_{1}}),
\end{equation}
where
\begin{align*}
& \theta _{1}^{*}+{{\theta }_{1}}=-2{{\tilde{b}}_{1}}\big[x+\big(4{{A}_{2}}{{{\tilde{a}}}_{1}}+12{{A}_{3}}\tilde{a}_{1}^{2}-32{{A}_{4}}\tilde{a}_{1}^{3}+32{{A}_{4}}{{{\tilde{a}}}_{1}}\tilde{b}_{1}^{2}-80{{A}_{5}}\tilde{a}_{1}^{4}
\\
&\quad\quad\quad\quad\ +160{{A}_{5}}\tilde{a}_{1}^{2}\tilde{b}_{1}^{2} +192{{A}_{6}}\tilde{a}_{1}^{5}-640{{A}_{6}}\tilde{a}_{1}^{3}\tilde{b}_{1}^{2}+192{{A}_{6}}{{{\tilde{a}}}_{1}}\tilde{b}_{1}^{4}+448{{A}_{7}}\tilde{a}_{1}^{6} \\
&\quad\quad\quad\quad\ -2240{{A}_{7}}\tilde{a}_{1}^{4}\tilde{b}_{1}^{2} +1344{{A}_{7}}\tilde{a}_{1}^{2}\tilde{b}_{1}^{4}-1024{{A}_{8}}\tilde{a}_{1}^{7}-7168{{A}_{8}}\tilde{b}_{1}^{4}\tilde{a}_{1}^{3}\\
&\quad\quad\quad\quad\ +7168{{A}_{8}}\tilde{b}_{1}^{2}\tilde{a}_{1}^{5}+1024{{A}_{8}}\tilde{b}_{1}^{6}{{{\tilde{a}}}_{1}}-4{{A}_{3}}\tilde{b}_{1}^{2}-16{{A}_{5}}\tilde{b}_{1}^{4}-64{{A}_{7}}\tilde{b}_{1}^{6}\big)t\big],\\&\theta _{1}^{*}-{\theta }_{1}=-2ix\tilde{a}_{{1}}+960itA_{{6}}{\tilde{a}_{{1}}^{4}}{\tilde{b}_{{1}}^{2}}+24itA_{{3}}\tilde{a}_{{1}}{\tilde{b}_{{1}}^{2}}+896itA_{{7}}\tilde{a}_{{1}}{\tilde{b}_{{1}}^{6}}+256itA_{{8}}{\tilde{a}_{{1}}^{8}}\\
&\quad\quad\quad\quad\ +16itA_{{4}}{\tilde{a}_{{1}}^{4}}+160itA_{{5}}\tilde{a}_{{1}}{\tilde{b}_{{1}}^{4}}-4itA_{{2
}}{\tilde{a}_{{1}}^{2}}+2688itA_{{7}}{\tilde{a}_{{1}}^{5}}{\tilde{b}_{{1}}^{2}}+16itA_{{4}}{\tilde{b}_{{1}}^{4}}\\
&\quad\quad\quad\quad\ -64itA_{{6}}{\tilde{a}_{{1}}^{6}}+4itA_{{2}}{\tilde{b}_{{1}}^{2}}-96itA_{{4}}{\tilde{a}_{{1}}^{2}}{\tilde{b}_{{1}}^{2}}-7168itA_{{8}}{\tilde{a}_{{1}}^{6}}{\tilde{b}_{
{1}}^{2}}-7168itA_{{8}}{\tilde{a}_{{1}}^{2}}{\tilde{b}_{{1}}^{6}}\\
&\quad\quad\quad\quad\ -4480itA_{{7}}{\tilde{a}_{{1}}^{3}}{\tilde{b}_{{1}}^{4}}+32itA_{{5}}{\tilde{a}_{{1}}^{5}}-8itA_{{3}}{\tilde{a}_{{1}}}^{3}-320itA_{{5}}{\tilde{a}_{{1}}^{3}}{\tilde{b}_{{1}}^{2}}-128itA_{{7}}{\tilde{a}_{{1}}^{7}}\\
&\quad\quad\quad\quad\ -960itA_{{6}}{\tilde{a}_{{1}}^{2}}{\tilde{b}_{{1}}
^{4}}+256itA_{{8}}{\tilde{b}_{{1}}^{8}}+64itA_{{6}}{\tilde{b}_{{1}}^{6}}+17920itA_{{8}}{\tilde{a}_{{1}}^{4}}{\tilde{b}_{{1}}^{4}}.
\end{align*}
Hence we can further write the bright one-soliton solution (25) as
\begin{equation}
\begin{aligned}
&q(x,t)=-2i\alpha _{1}^{*}{{\tilde{b}}_{1}}{{\text{e}}^{-{{\xi }_{1}}}}{{\text{e}}^{\theta _{1}^{*}-{{\theta }_{1}}}}\text{sech}\big\{-2{{\tilde{b}}_{1}}\big[x+\big(4{{A}_{2}}{{{\tilde{a}}}_{1}}+12{{A}_{3}}\tilde{a}_{1}^{2}-32{{A}_{4}}\tilde{a}_{1}^{3}\\
&\quad\quad\quad\quad+32{{A}_{4}}{{{\tilde{a}}}_{1}}\tilde{b}_{1}^{2}-80{{A}_{5}}\tilde{a}_{1}^{4}
+160{{A}_{5}}\tilde{a}_{1}^{2}\tilde{b}_{1}^{2} +192{{A}_{6}}\tilde{a}_{1}^{5}-640{{A}_{6}}\tilde{a}_{1}^{3}\tilde{b}_{1}^{2}\\
&\quad\quad\quad\quad+192{{A}_{6}}{{{\tilde{a}}}_{1}}\tilde{b}_{1}^{4}+448{{A}_{7}}\tilde{a}_{1}^{6}-2240{{A}_{7}}\tilde{a}_{1}^{4}\tilde{b}_{1}^{2} +1344{{A}_{7}}\tilde{a}_{1}^{2}\tilde{b}_{1}^{4}-1024{{A}_{8}}\tilde{a}_{1}^{7}\\
&\quad\quad\quad\quad-7168{{A}_{8}}\tilde{b}_{1}^{4}\tilde{a}_{1}^{3} +7168{{A}_{8}}\tilde{b}_{1}^{2}\tilde{a}_{1}^{5}+1024{{A}_{8}}\tilde{b}_{1}^{6}{{{\tilde{a}}}_{1}}-4{{A}_{3}}\tilde{b}_{1}^{2}-16{{A}_{5}}\tilde{b}_{1}^{4}\\
&\quad\quad\quad\quad-64{{A}_{7}}\tilde{b}_{1}^{6}\big)t\big]+{{\xi }_{1}}\big\},
\end{aligned}
\end{equation}
from which it is indicated that the solution (26) takes the shape of hyperbolic secant function with peak amplitude
\[\mathcal{H}=2\left| \alpha _{1}^{*} \right|{{\tilde{b}}_{1}}{{\text{e}}^{-{{\xi }_{1}}}}\]
and velocity
\begin{equation*}
\begin{aligned}
&\mathcal{V}=-4{{A}_{2}}{{{\tilde{a}}}_{1}}-12{{A}_{3}}\tilde{a}_{1}^{2}+32{{A}_{4}}\tilde{a}_{1}^{3}-32{{A}_{4}}{{{\tilde{a}}}_{1}}\tilde{b}_{1}^{2}+80{{A}_{5}}\tilde{a}_{1}^{4}
-160{{A}_{5}}\tilde{a}_{1}^{2}\tilde{b}_{1}^{2}\\
&\quad\quad-192{{A}_{6}}\tilde{a}_{1}^{5}+640{{A}_{6}}\tilde{a}_{1}^{3}\tilde{b}_{1}^{2}-192{{A}_{6}}{{{\tilde{a}}}_{1}}\tilde{b}_{1}^{4}-448{{A}_{7}}\tilde{a}_{1}^{6}+2240{{A}_{7}}\tilde{a}_{1}^{4}\tilde{b}_{1}^{2} \\&\quad\quad-1344{{A}_{7}}\tilde{a}_{1}^{2}\tilde{b}_{1}^{4}+1024{{A}_{8}}\tilde{a}_{1}^{7}+7168{{A}_{8}}\tilde{b}_{1}^{4}\tilde{a}_{1}^{3} -7168{{A}_{8}}\tilde{b}_{1}^{2}\tilde{a}_{1}^{5}\\
&\quad\quad-1024{{A}_{8}}\tilde{b}_{1}^{6}{{{\tilde{a}}}_{1}}+4{{A}_{3}}\tilde{b}_{1}^{2}+16{{A}_{5}}\tilde{b}_{1}^{4}+64{{A}_{7}}\tilde{b}_{1}^{6}.
\end{aligned}
\end{equation*}

To show the localized structures and dynamic behaviors of one-soliton solution (26), we select the involved parameters as ${{\tilde{a}}}_{1}=0.3,{{\tilde{b}}}_{1}=0.2,\alpha_{1}=A_{2}=A_{3}=A_{4}=A_{5}=A_{6}=A_{7}=A_{8}=1,\xi_{1}=0.$
The plots are depicted in Figures 1--3.

Then for the case of $N=2$, the bright two-soliton solution for Equation (1) is generated as
\begin{equation}\begin{aligned}
&q(x,t)=\frac{2}{{{M}_{12}}{{M}_{21}}-{{M}_{11}}{{M}_{22}}}\big(\alpha _{1}^{*}{{\beta }_{1}}{{\text{e}}^{-{{\theta }_{1}}+\theta _{1}^{*}}}{{M}_{22}}-\alpha _{2}^{*}{{\beta }_{1}}{{\text{e}}^{-{{\theta }_{1}}+\theta _{2}^{*}}}{{M}_{12}}\\&\quad\quad\quad\quad-\alpha _{1}^{*}{{\beta }_{2}}{{\text{e}}^{-{{\theta }_{2}}+\theta _{1}^{*}}}{{M}_{21}}+\alpha _{2}^{*}{{\beta }_{2}}{{\text{e}}^{-{{\theta }_{2}}+\theta _{2}^{*}}}{{M}_{11}}\big),
\end{aligned}\end{equation}
where
\[\begin{aligned}
 & {{M}_{11}}=\frac{{{\left| {{\alpha }_{1}} \right|}^{2}}{{\text{e}}^{\theta _{1}^{*}+{{\theta }_{1}}}}+{{\left| {{\beta }_{1}} \right|}^{2}}{{\text{e}}^{-\theta _{1}^{*}-{{\theta }_{1}}}}}{{{\varsigma }_{1}}-\varsigma _{1}^{*}},\quad{{M}_{12}}=\frac{\alpha _{1}^{*}{{\alpha }_{2}}{{\text{e}}^{\theta _{1}^{*}+{{\theta }_{2}}}}+\beta _{1}^{*}{{\beta }_{2}}{{\text{e}}^{-\theta _{1}^{*}-{{\theta }_{2}}}}}{{{\varsigma }_{2}}-\varsigma _{1}^{*}}, \\
 & {{M}_{21}}=\frac{\alpha _{2}^{*}{{\alpha }_{1}}{{\text{e}}^{\theta _{2}^{*}+{{\theta }_{1}}}}+\beta _{2}^{*}{{\beta }_{1}}{{\text{e}}^{-\theta _{2}^{*}-{{\theta }_{1}}}}}{{{\varsigma }_{1}}-\varsigma _{2}^{*}},\quad{{M}_{22}}=\frac{{{\left| {{\alpha }_{2}} \right|}^{2}}{{\text{e}}^{\theta _{2}^{*}+{{\theta }_{2}}}}+{{\left| {{\beta }_{2}} \right|}^{2}}{{\text{e}}^{-\theta _{2}^{*}-{{\theta }_{2}}}}}{{{\varsigma }_{2}}-\varsigma _{2}^{*}},\\
 &\theta_{1}=i{{\varsigma }_{1}}x+i\big(2{{A}_{2}}\varsigma _{1}^{2}+4{{A}_{3}}\varsigma _{1}^{3}-8{{A}_{4}}\varsigma _{1}^{4}-16{{A}_{5}}\varsigma _{1}^{5}+32{{A}_{6}}\varsigma _{1}^{6}+64{{A}_{7}}\varsigma _{1}^{7}-128{{A}_{8}}\varsigma _{1}^{8}\big)t,\\
 &\theta_{2}=i{{\varsigma }_{2}}x+i\big(2{{A}_{2}}\varsigma _{2}^{2}+4{{A}_{3}}\varsigma _{2}^{3}-8{{A}_{4}}\varsigma _{2}^{4}-16{{A}_{5}}\varsigma _{2}^{5}+32{{A}_{6}}\varsigma _{2}^{6}+64{{A}_{7}}\varsigma _{2}^{7}-128{{A}_{8}}\varsigma _{2}^{8}\big)t,
\end{aligned}\]
and ${{\varsigma}_{1}}={{\tilde{a}}_{1}}+i{{\tilde{b}}_{1}},{{\varsigma}_{2}}={{\tilde{a}}_{2}}+i{{\tilde{b}}_{2}}$.

After assuming that ${{\beta }_{1}}={{\beta }_{2}}=1$ and ${{\alpha }_{1}}={{\alpha }_{2}}$ as well as ${{\left| {{\alpha }_{1}} \right|}^{2}}={{\text{e}}^{2{{\xi }_{1}}}}$, the bright two-soliton solution (27) becomes
\begin{equation}
q(x,t)=\frac{2}{{{M}_{12}}{{M}_{21}}-{{M}_{11}}{{M}_{22}}}\big(\alpha _{1}^{*}{{\text{e}}^{-{{\theta }_{1}}+\theta _{1}^{*}}}{{M}_{22}}-\alpha _{2}^{*}{{\text{e}}^{-{{\theta }_{1}}+\theta _{2}^{*}}}{{M}_{12}}-\alpha _{1}^{*}{{\text{e}}^{-{{\theta }_{2}}+\theta _{1}^{*}}}{{M}_{21}}+\alpha _{2}^{*}{{\text{e}}^{-{{\theta }_{2}}+\theta _{2}^{*}}}{{M}_{11}}\big),
\end{equation}
where
\[\begin{aligned}
 & {{M}_{11}}=-\frac{i}{{{{\tilde{b}}}_{1}}}{{\text{e}}^{{{\xi }_{1}}}}\cosh (\theta _{1}^{*}+{{\theta }_{1}}+{{\xi }_{1}}), \\
 & {{M}_{12}}=\frac{2{{\text{e}}^{{{\xi }_{1}}}}}{({{{\tilde{a}}}_{2}}-{{{\tilde{a}}}_{1}})+i({{{\tilde{b}}}_{1}}+{{{\tilde{b}}}_{2}})}\cosh (\theta _{1}^{*}+{{\theta }_{2}}+{{\xi }_{1}}), \\
 & {{M}_{21}}=\frac{2{{\text{e}}^{{{\xi }_{2}}}}}{({{{\tilde{a}}}_{1}}-{{{\tilde{a}}}_{2}})+i({{{\tilde{b}}}_{1}}+{{{\tilde{b}}}_{2}})}\cosh (\theta _{2}^{*}+{{\theta }_{1}}+{{\xi }_{2}}), \\
 & {{M}_{22}}=-\frac{i}{{{{\tilde{b}}}_{2}}}{{\text{e}}^{{{\xi }_{2}}}}\cosh (\theta _{2}^{*}+{{\theta }_{2}}+{{\xi }_{2}}). \\
\end{aligned}\]

The localized structures and dynamic behaviors of two-soliton solution (28) are depicted in Figure 4 via
a selection of the parameters as ${{\tilde{a}}}_{1}=0.3,{{\tilde{b}}}_{1}=0.1,{{\tilde{b}}}_{2}=0.2,\alpha_{1}=\alpha_{2}=A_{2}=A_{3}=A_{4}=A_{5}=A_{6}=A_{7}=A_{8}=1,{{\tilde{a}}}_{2}=\xi_{1}=\xi_{2}=0.$

\begin{figure}
\begin{center}
\subfigure[]{\resizebox{0.38\hsize}{!}{\includegraphics*{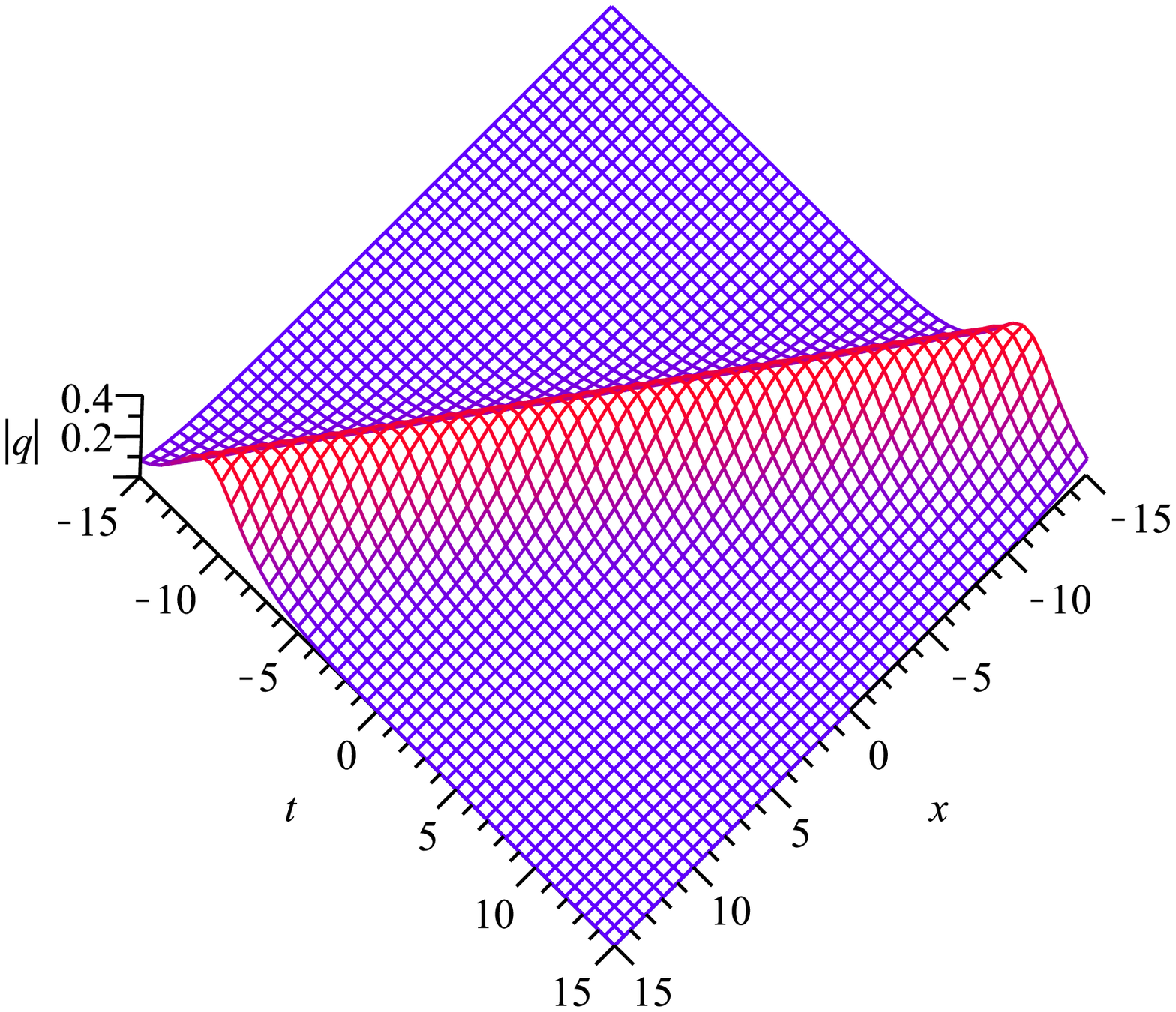}}}
\subfigure[]{\resizebox{0.38\hsize}{!}{\includegraphics*{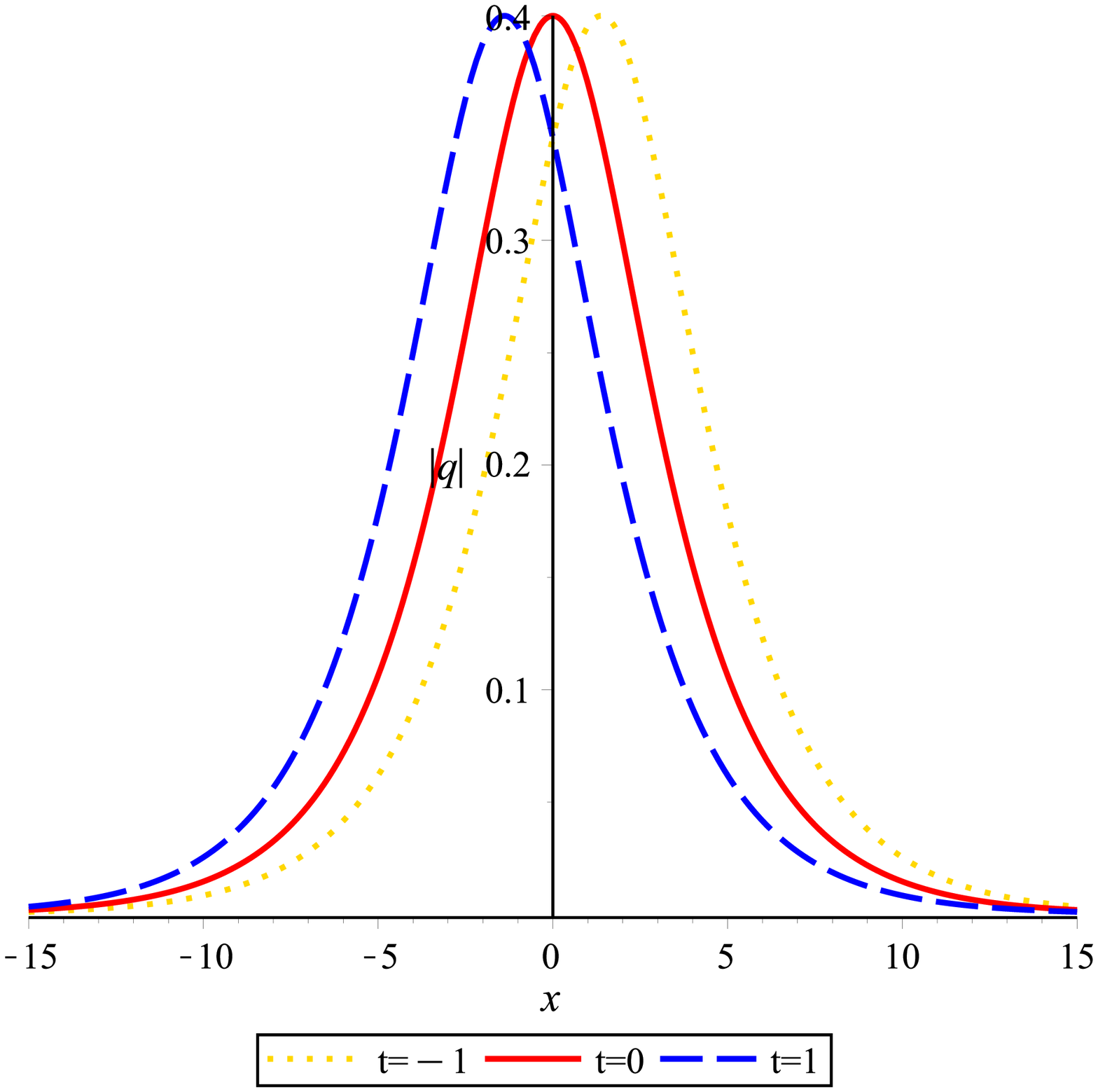}}}
\parbox[c]{13.0cm}{\footnotesize
{\bf Figure 1.}~Plots of one-soliton solution (26): (a) Perspective view of modulus of $q$; (b) The soliton along the $x$-axis with different time in Figure~1(a).}
\end{center}
\addtocounter{subfigure}{-2}
\end{figure}

\begin{figure}
\begin{center}
\subfigure[]{\resizebox{0.38\hsize}{!}{\includegraphics*{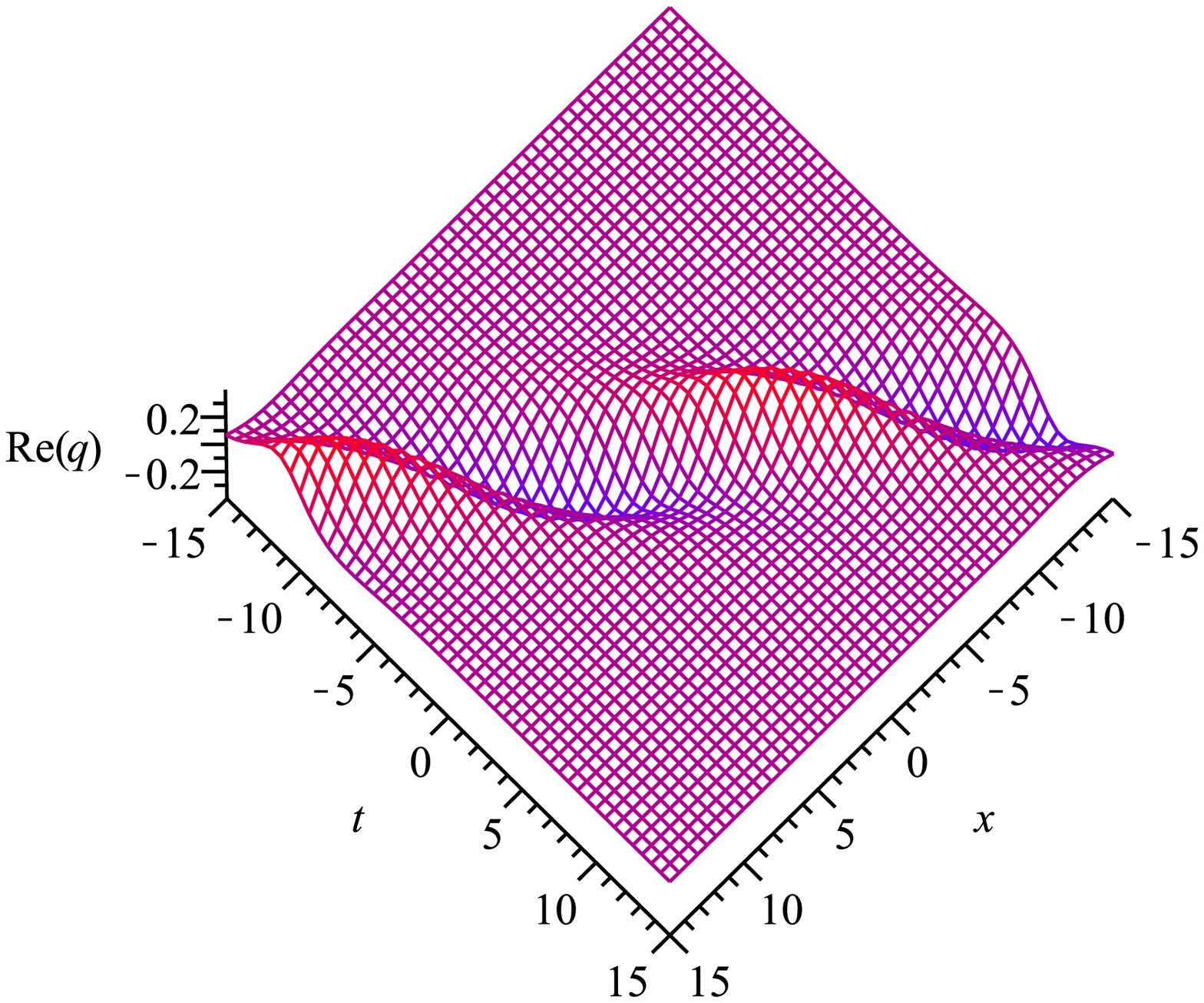}}}
\subfigure[]{\resizebox{0.38\hsize}{!}{\includegraphics*{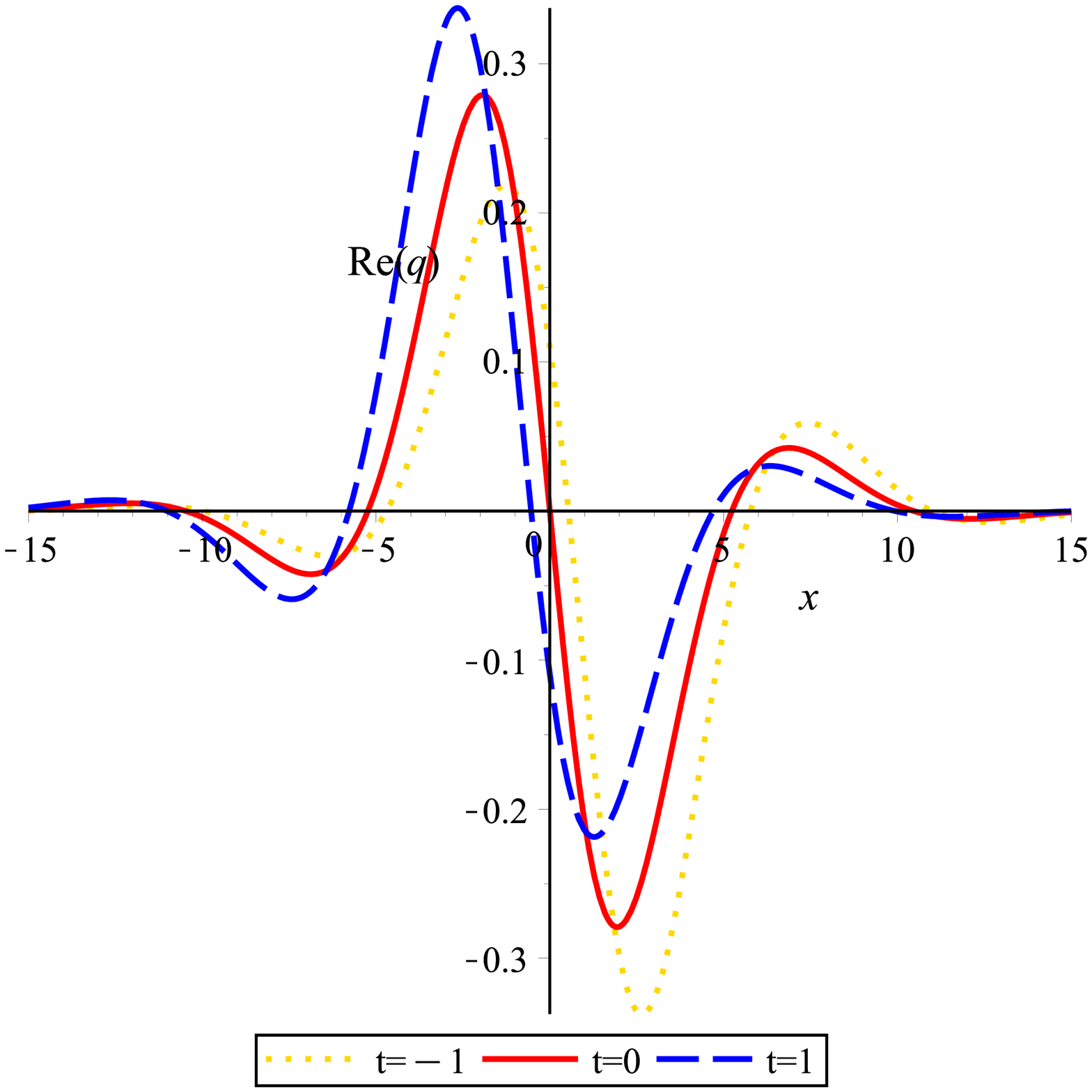}}}
\parbox[c]{13.0cm}{\footnotesize
{\bf Figure 2.}~Plots of one-soliton solution (26): (a) Perspective view of real part of $q$; (b) The soliton along the $x$-axis with different time in Figure~2(a).}
\end{center}
\addtocounter{subfigure}{-2}
\end{figure}

\begin{figure}
\begin{center}
\subfigure[]{\resizebox{0.38\hsize}{!}{\includegraphics*{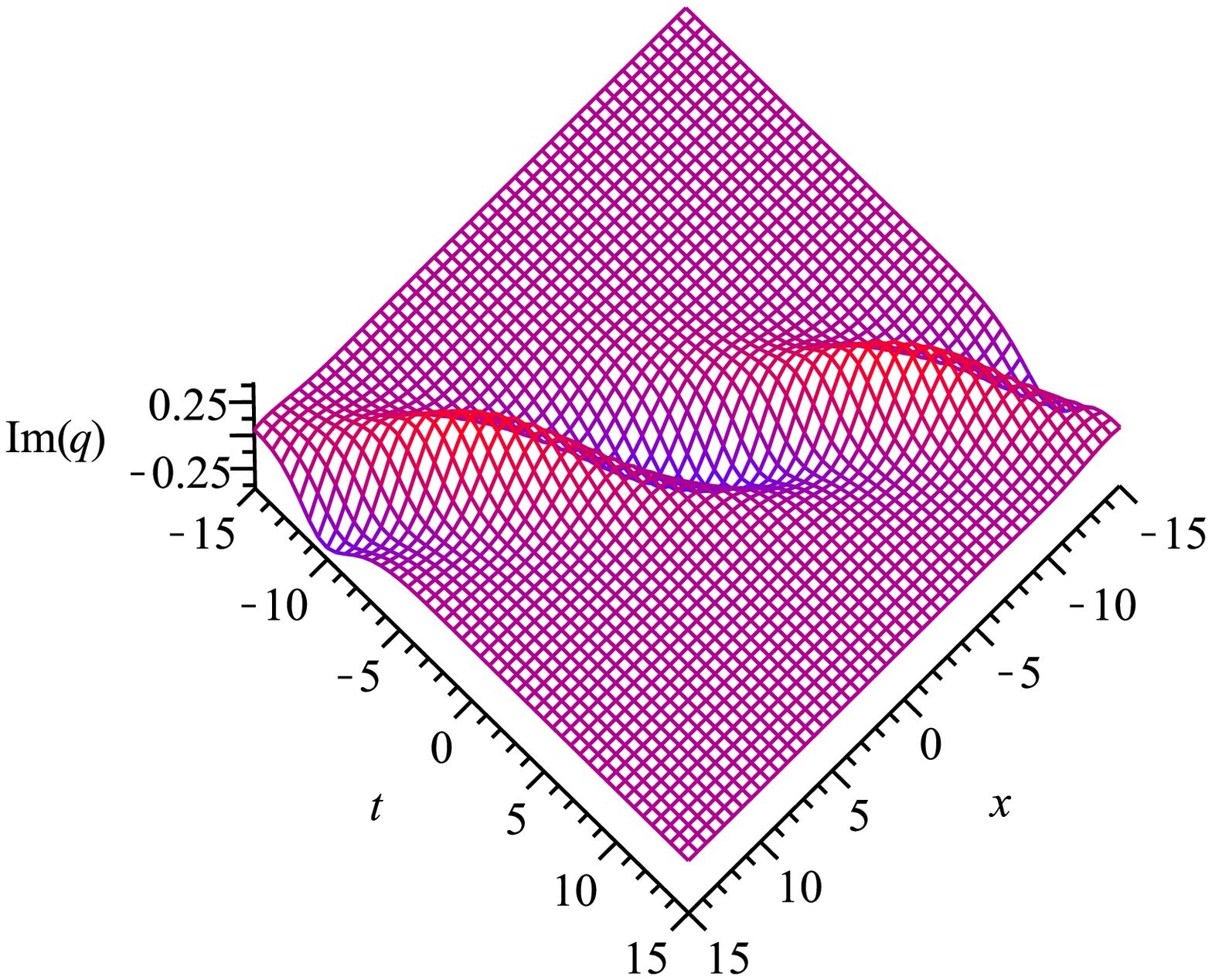}}}
\subfigure[]{\resizebox{0.38\hsize}{!}{\includegraphics*{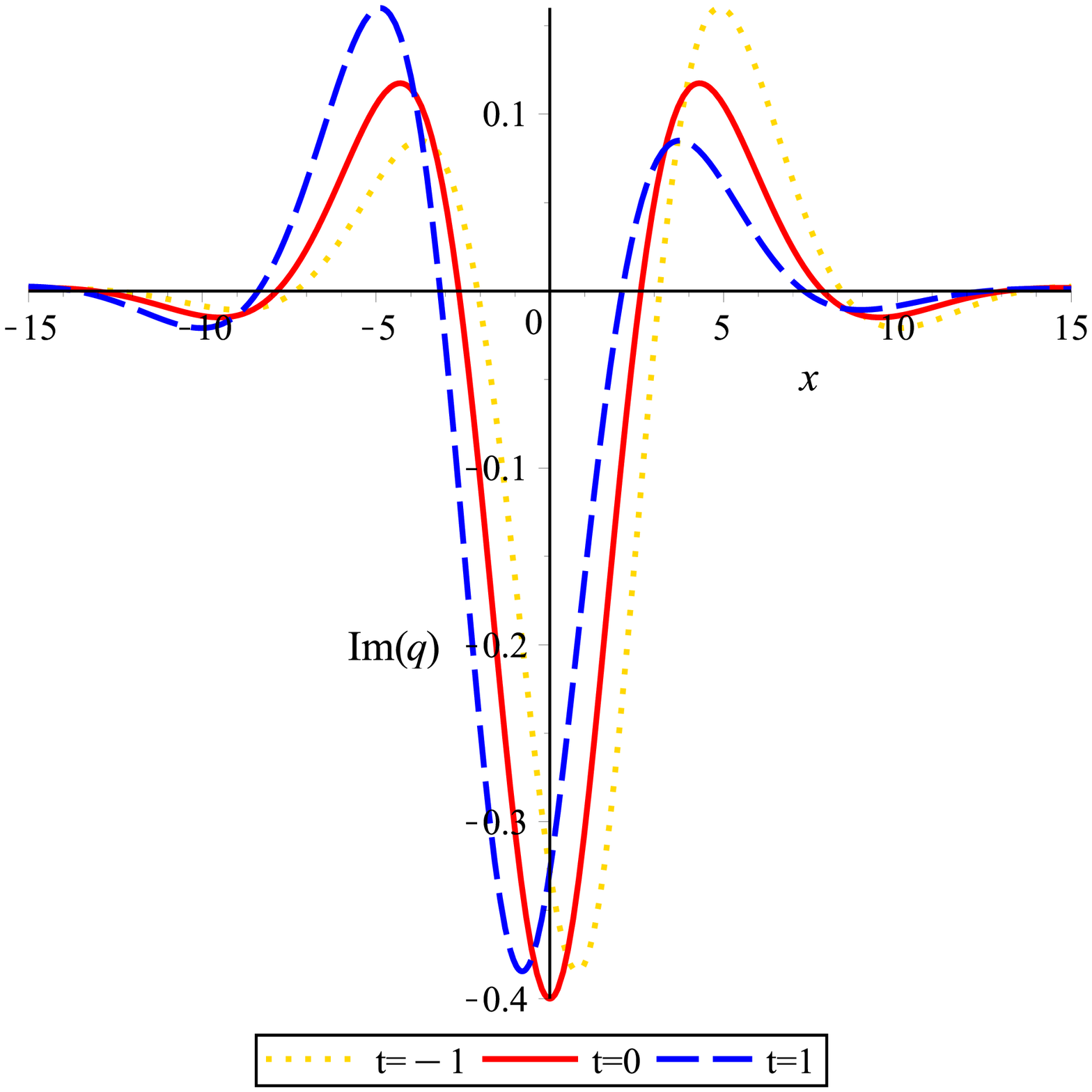}}}
\parbox[c]{13.0cm}{\footnotesize
{\bf Figure 3.}~Plots of one-soliton solution (26): (a) Perspective view of imaginary part of $q$; (b) The soliton along the $x$-axis with different time in Figure~3(a).}
\end{center}
\addtocounter{subfigure}{-2}
\end{figure}

\begin{figure}
\begin{center}
\subfigure[]{\resizebox{0.38\hsize}{!}{\includegraphics*{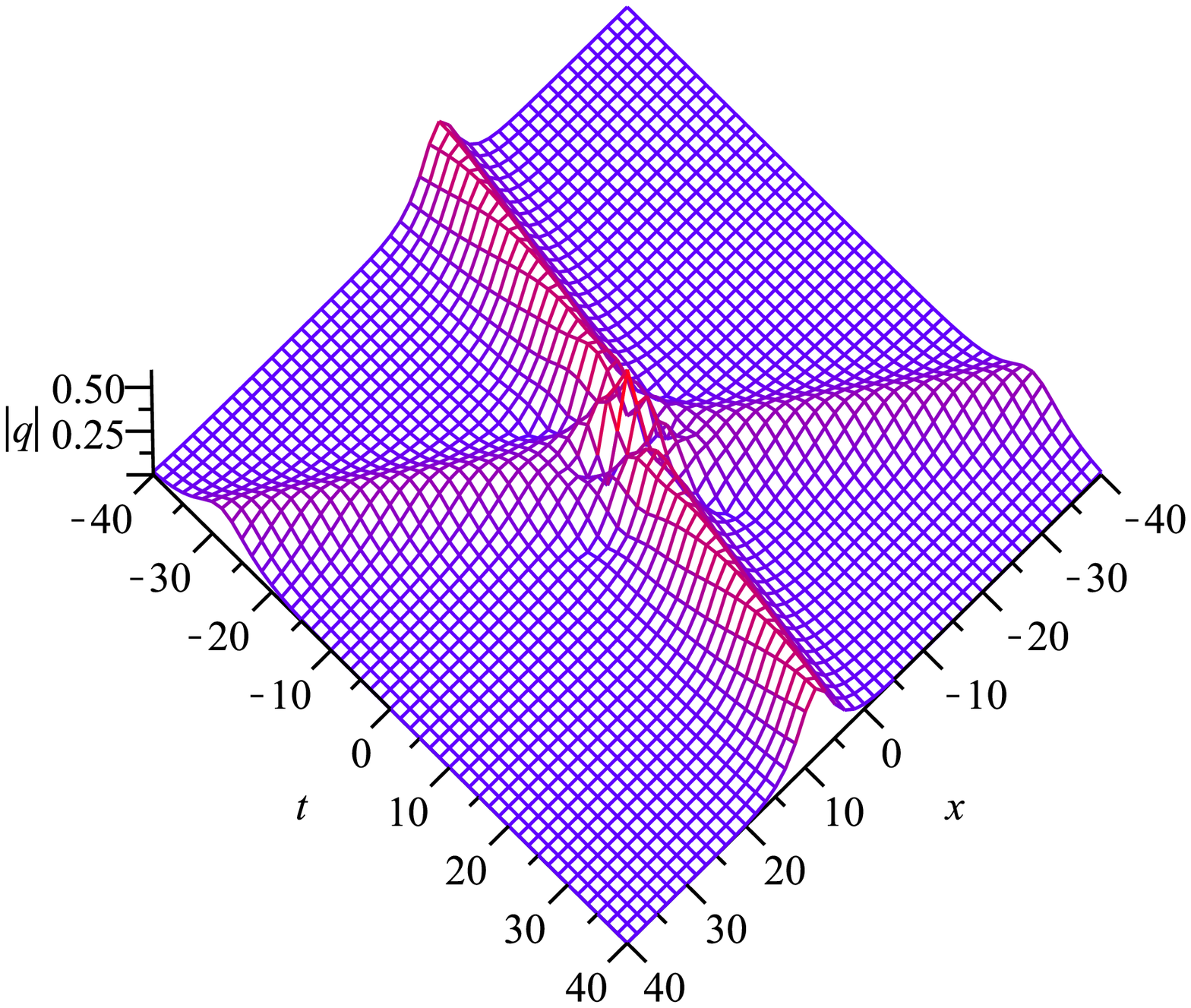}}}
\subfigure[]{\resizebox{0.38\hsize}{!}{\includegraphics*{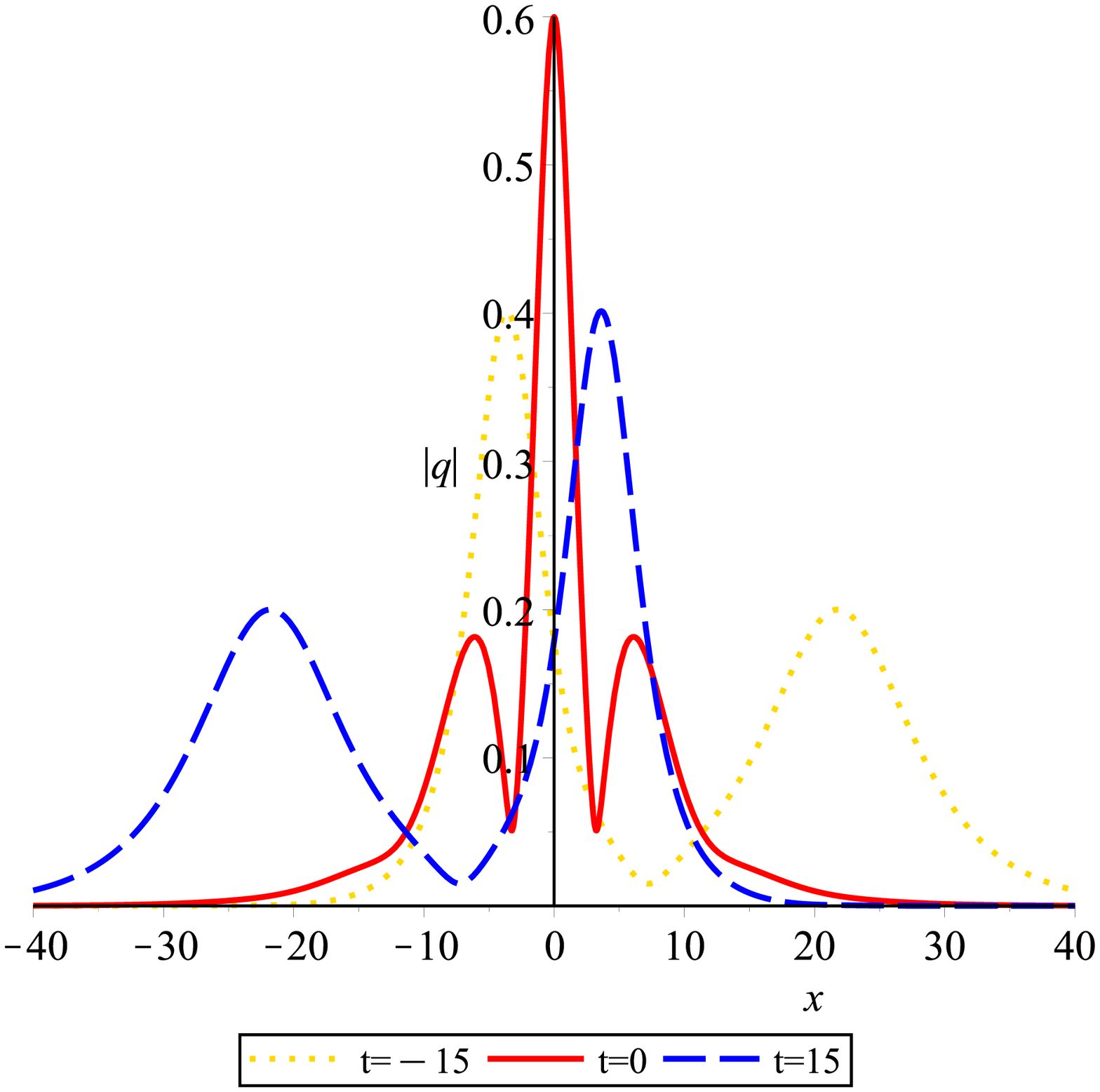}}}
\parbox[c]{13.0cm}{\footnotesize
{\bf Figure 4.}~Plots of two-soliton solution (28): (a) Perspective view of modulus of $q$; (b) The soliton along the $x$-axis with different time in Figure~4(a).}
\end{center}
\end{figure}

\section{Conclusion}
In this investigation, the aim was to explore multi-soliton solutions for an eighth-order nonlinear Schr\"{o}dinger equation arising in an optical fiber. The method we resort to was the Riemann-Hilbert approach which is based on a Riemann-Hilbert problem. Therefore, we first described a Riemann-Hilbert problem via analyzing the spectral problem of the Lax pair. After solving the obtained Riemann-Hilbert problem corresponding to the reflectionless case, we finally generated the expression of general $N$-soliton solution to the eighth-order nonlinear Schr\"{o}dinger equation. In addition, the localized structures and dynamic behaviors of bright one- and two-soliton solutions were shown graphically via suitable choices of the involved parameters.

\section*{Funding}
This work was supported by the National Natural Science Foundation of China (Grant Nos. 61072147 and 11271008).



\end{document}